\documentclass{pasa2}%

\usepackage{graphicx}
\usepackage{xcolor}
\definecolor{dodgerblue}{RGB}{30, 144, 255}

\usepackage{aas_macros}
\usepackage{hyperref} 
\hypersetup{colorlinks,citecolor=dodgerblue,linkcolor=dodgerblue,urlcolor=dodgerblue}

\usepackage{amsmath}

\usepackage{booktabs}
\usepackage{threeparttable}
\usepackage{subcaption,cleveref}
\usepackage{cprotect}
\usepackage{multirow}
\usepackage{sansmath}
\AtBeginEnvironment{tablenotes}{\sffamily\sansmath}

\usepackage{times}
\usepackage{mathptmx}  
\usepackage{tgheros}   
\usepackage{beramono}  

\newcommand{\tabft}[1]{(#1)}
\newcounter{ft}
\newcommand{\ft}[1]{\noindent%
  ~\refstepcounter{ft}\tabft{\alph{ft}\label{#1}}}
  
\makeatletter\ifdefined\Hy@StartlinkName
\def\addOneNestingLevelStartLink{%
  \gdef\Hy@StartlinkName##1##2{%
    \sbox0{\Hy@StartlinkNameOrig{##1}{##2}}\usebox0
    \global\let\Hy@StartlinkName\Hy@StartlinkNameOrig%
  }%
}
\def\addOneNestingLevelEndLink{%
  \gdef\pdfendlink{%
    \sbox0{\pdfendlinkOrig}\usebox0%
    \global\let\pdfendlink\pdfendlinkOrig%
  }%
}
\let\Hy@StartlinkNameOrig\Hy@StartlinkName
\let\pdfendlinkOrig\pdfendlink
\else
\let\addOneNestingLevelStartLink\relax
\let\addOneNestingLevelEndLink\relax
\fi
\makeatother

\def\degr{\ensuremath{^\circ}}
\def\arcmin{\ensuremath{^{\prime}}}
\def\arcsec{\ensuremath{^{\prime\prime}}}

\def\xmm{XMM-\textit{Newton}}

\newcommand{\CORRS}[1]{{\color{black} #1}}
\newcommand{\corrs}[1]{{\color{black} #1}}
\newcommand{\CORRStwo}[1]{{\color{black} #1}}
\newcommand{\CORRSthree}[1]{{\color{black} #1}}
\newcommand{\CORRSfour}[1]{{\color{black} #1}}

\def\alphaNonly{\corrs{\ensuremath{-1.18 \pm 0.10}}}
\def\alphaNEonly{\corrs{\ensuremath{ -1.54 \pm 0.45}}}
\def\alphaSonly{\corrs{\ensuremath{ -1.52 \pm 0.10}}}
\def\alphaN{\ensuremath{\corrs{\alpha_\mathrm{NW,full} = -1.18 \pm 0.10}}}
\def\alphaNE{\ensuremath{\corrs{\alpha_\mathrm{NW} = -1.54 \pm 0.45}}}
\def\alphaS{\ensuremath{\corrs{\alpha_\mathrm{SE} = -1.52 \pm 0.10}}}

\def\powerS{\ensuremath{\corrs{14.0 \pm 1.6}}}
\def\powerN{\ensuremath{\corrs{9.1 \pm 1.0}}}
\def\powerW{\ensuremath{\corrs{6.2 \pm 1.8}}}

\title[SPT-CL J2032$-$5627: a new double relic cluster]{SPT-CL~J2032$-$5627: a new Southern double relic cluster observed with ASKAP}
\author[S.~W.~Duchesne et al.]{S.~W. Duchesne$^1$\thanks{email: \url{stefan.duchesne.astro@gmail.com}},  M. Johnston-Hollitt$^{1,2}$, I. Bartalucci$^{3,4}$, T. Hodgson$^1$, and G.~W. Pratt$^{3}$
\affil{$^1$International Centre for Radio Astronomy Research (ICRAR), Curtin University, Bentley, WA 6102, Australia}%
\affil{$^2$Curtin Institute for Computation, Curtin University,
GPO Box U1987, Perth, WA 6845, Australia}%
\affil{$^3$AIM, CEA, CNRS, Universit\'{e} Paris-Saclay, Universit\'{e} Paris Diderot, Sorbonne Paris Cit\'{e}, F-91191 Gif-sur-Yvette, France}
\affil{$^4$INAF - Istituto di Astrofisica Spaziale e Fisica Cosmica di Milano, Via A. Corti 12, 20133 Milano, Italy}
}%

\jid{PASA}
\doi{10.1017/pas.\the\year.xxx}
\jyear{\the\year}
\documentstatus{\textcolor{green}{accepted} for publication in PASA}

\def\spt{SPT-CL~J2032$-$5627}
\def\jybeam{\ensuremath{\mathrm{Jy}\,\mathrm{beam}^{-1}}}

\begin{document}

\begin{frontmatter}
\maketitle
\rule{\linewidth}{0.75pt}\vspace{11.5pt}
\begin{abstract}
We present a radio and X-ray analysis of the galaxy cluster \spt. Investigation of public data from the Australian Square Kilometre Array Pathfinder (ASKAP) at 943~MHz shows two previously undetected radio relics at either side of the cluster. For both relic sources we utilise archival Australia Telescope Compact Array (ATCA) data at 5.5~GHz in conjunction with the new ASKAP data to determine that both have steep integrated radio spectra (\alphaS~and \alphaN~for the southeast and northwest relic sources, respectively). No shock is seen in \xmm\ observations, however, the southeast relic is preceded by a cold front in the X-ray--emitting intra-cluster medium. We suggest the lack of \CORRS{a} detectable shock may be due to instrumental limitations, comparing the situation to the southeast relic in Abell~3667. We compare the relics to the population of double relic sources and find they are located below the current power--mass ($P$--$M$) scaling relation. We present an analysis of the low-surface brightness sensitivity of ASKAP and the ATCA, the excellent sensitivity of both allow the ability to find heretofore undetected diffuse sources, suggesting these low-power radio relics will become more prevalent in upcoming large-area radio surveys such as the Evolutionary Map of the Universe (EMU). 
\end{abstract}

\rule{\linewidth}{0.75pt}\vspace{11.5pt}
\end{frontmatter}

\section{Introduction}\label{sec:intro}
Clusters of galaxies are the largest virialized systems in the Universe \citep[e.g.][]{pee80,oort_superclusters_1983}, and can best be described as gravitational potential wells comprised predominantly of dark matter \citep[e.g.][]{blumenthal_formation_1984}, hosting tens to thousands of optically luminous galaxies embedded in an X-ray emitting plasma. Galaxy clusters form hierarchically  through accretion and highly energetic merger events, and in the $\Lambda$ Cold Dark Matter ($\Lambda$CDM) cosmology, energy is provided to the intra-cluster medium (ICM) through the infall of ICM gas into the potential well of the cluster dark matter halo \citep[see e.g.][for a review]{kravtsov_formation_2012}. The energy provided to the ICM creates shocks, adiabatic compression, and turbulence within the ICM, as well as providing additional thermal energy to the ICM gas. \par
Within predominantly unrelaxed (e.g. merging) galaxy clusters, large-scale ($\sim 1$~Mpc) radio synchrotron emission has been observed, often either coincident with the centrally located X-ray--emitting ICM plasma (so-called \textit{giant radio halos}, e.g. in the Coma Cluster; \citealt{wil70}, or the halo in Abell~2744; \citealt{gfg+01}), or peripherally located and co-spatially with X-ray--detected shocks (\textit{radio relics} or \textit{radio shocks}, e.g. the double relic system in Abell~3667; \citealt{mj-h,Finoguenov2010}). These sources are characterised by not only their large spatial scales, but also their steep, approximately power law synchrotron spectra, with a lack of obvious host galaxy \citep[for a review see][]{vda+19}. Additionally, smaller-scale ($\lesssim 400$~kpc), steep-spectrum radio emission is seen within clusters classified as either \textit{mini-halos} \CORRS{\citep[see e.g.][]{Giacintucci2017,Giacintucci2019}} or \textit{radio phoenices} \CORRS{\citep[see e.g.][and see also \citealt{Ensslin2001}]{slee2001}} which have some observational similarities \CORRS{(e.g. steep spectrum, low surface brightness)} to their larger cousins which can make them difficult to differentiate in some cases.

Such large-scale cluster emission is thought to be generated via in situ acceleration (or \textit{re-}acceleration) of particles. In the case of radio relics (or radio shocks), the (re-)acceleration process is thought to be primarily diffusive-shock acceleration (DSA; see e.g. \citealt{Axford1977,bell_acceleration_1978a,bell_acceleration_1978b,Blandford1978}; and, e.g. \citealt{je91}). DSA is largely consistent with the observed radio properties, and while a lack of observed $\gamma$-rays seen co-spatially with radio relics challenges simulations of this process, \CORRS{under certain shock and environmental conditions, e.g. specific shock orientations \citep{wittor_testing_2017}, lower (re-)acceleration efficiency \citep{Vazza2016}, or injection of fossil electrons from cluster radio galaxies \citep{Pinzke2013}, DSA may still be invoked}. If radio relics are truly generated from shocks in the ICM, then we may consider the observed double relic systems, hosting two relics on opposites sides of the host cluster, a suitable laboratory to explore the underlying physics of cluster mergers and the resultant emission. In double relic systems, the merger is thought to be not only between two significant subclusters, but also the merger axis is likely to be close to the plane of sky so projection-related effects (e.g. relic size, location, and brightness) are minimised and \CORRS{these} important physical parameters can be more accurately derived \CORRS{providing a more complete understanding of the relationships between radio relics and their host clusters} \citep[e.g.][]{mj-h08,Bonafede2012,deGasperin2014}.\par

\subsection{\spt}\label{sec:intro:spt}
\begin{figure*}[t]
    \centering
    \includegraphics[width=1\linewidth]{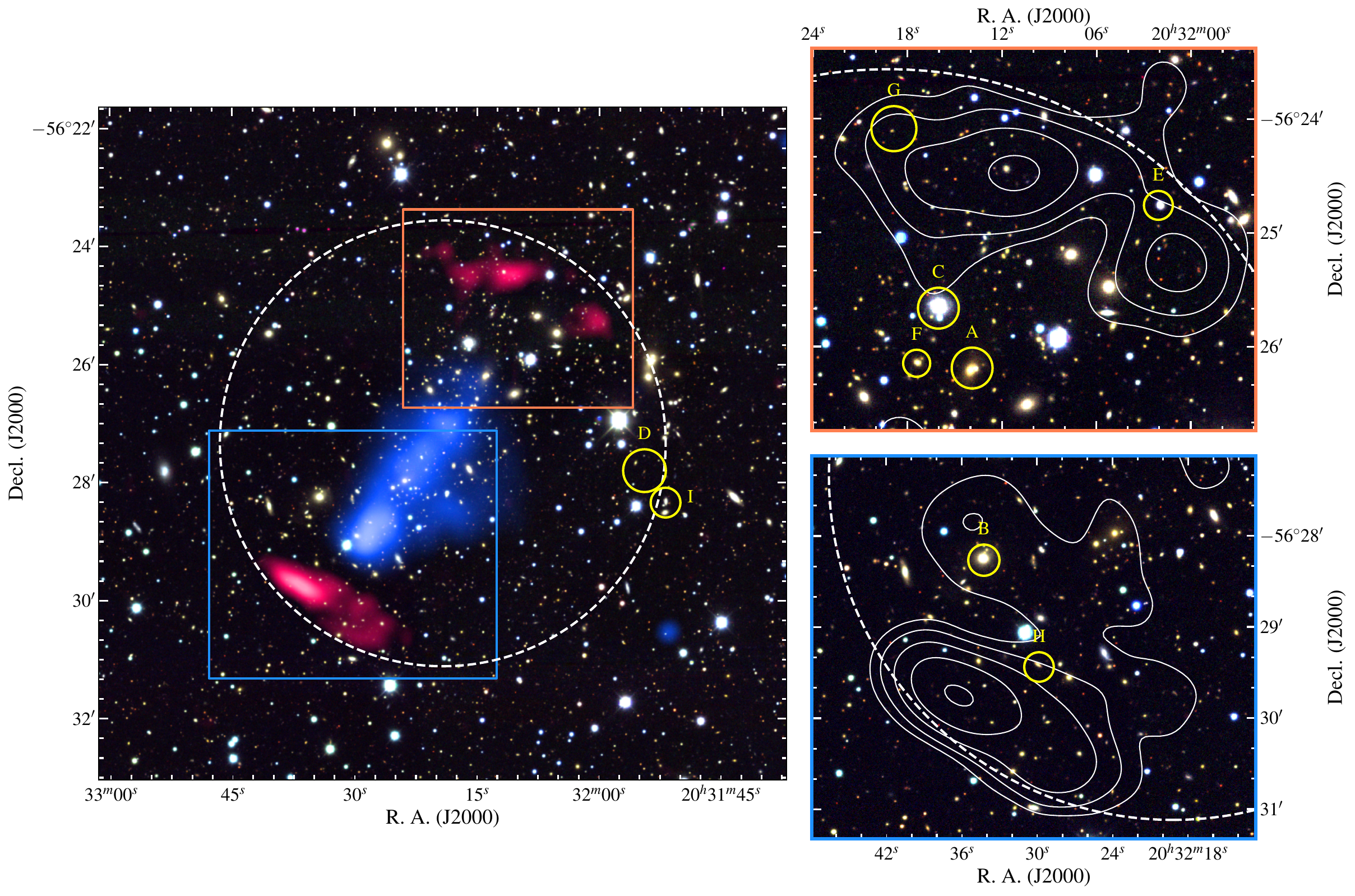}
    \caption{\label{fig:rgb} \spt: The background is an RGB image make using the $i$, $r$, and $g$ bands of the Dark Energy Survey Data Release 1 \citep[DES DR1;][]{des1,des2,decam}. In the left panel 0.943~GHz ASKAP robust $+0.25$ source-subtracted data are shown (magenta, \CORRS{$16.2^{\prime\prime}$ resolution}) overlaid with \xmm\ data (blue). The side panels feature discrete-source--subtracted, low resolution \CORRS{($36^{\prime\prime}$)} ASKAP data, starting at $3\sigma_\mathrm{rms}$ \CORRS{($\sigma_\text{rms} = 54$~$\mu$Jy\,beam$^{-1}$)} and increasing with factors of 2. The white, dashed circle is centered on \spt~and has a radius of 1~Mpc. \CORRS{The labelled objects are the main contaminating sources and} are discussed in the text, but are not an exhaustive list of subtracted discrete sources. The locations of the right panels are shown in the left panel with blue (bottom, southeast relic) and orange (top, northwest relic) boxes. ASKAP and \xmm\ data are described in Sections \ref{sec:data:askap} and \ref{sec:data:xmm}, respectively.}
\end{figure*}

\spt~is a galaxy cluster detected with the South Pole Telescope (SPT) as part of the SPT Cluster survey using the Sunyaev--Zel'dovich (SZ) effect to identify massive, distant clusters \citep{Song2012}. \citet{Song2012} report a redshift of $z=0.284$ via spectroscopy of cluster members. The corresponding \textit{Planck} catalogue of SZ sources \citep{planck15} reports an SZ-derived mass of $M_\mathrm{SZ} = 5.74^{+0.56}_{-0.59}\times10^{14}$~M$_\odot$. The cluster had previously been detected via X-ray (RXC~J2032.1$-$5627; \citealp{REFLEX}) \CORRS{and had been cross-matched with Abell~3685 (at $z=0.062$; \citealt{sr99}). We suggest that the redshift associated with Abell~3685 is derived from an isolated foreground galaxy (2MASS~J20321605$-$5625390, indicated on Fig. \ref{fig:rgb} as ``C''). It is unclear whether Abell~3685 is a distinct foreground cluster or if there is only a single cluster a $z=0.284$ along the line of sight.} \citet{Bulbul2019} report an X-ray--derived mass of $M_\mathrm{X,500} = 4.77^{+0.71}_{-0.63}\times10^{14}$~M$_\odot$ for \spt, consistent with the SZ-derived mass. \par
In this work we report on the detection of two heretofore unknown radio relic sources within \spt. A number of Australian radio telescopes have covered the cluster in a combination of surveys and pointed observations, namely: the Australia Telescope Compact Array \citep[ATCA;][]{fbw92}, the Molonglo Observatory Synthesis Telescope \citep[MOST;][]{Mills1981}, the Murchison Widefield Array \citep[MWA;][]{tgb+13,wtt+18}, and the Australian Square Kilometre Array Pathfinder \citep[ASKAP;][]{askap1,askap2,askap3}. Along with the radio data, archival \xmm\ data are available. This paper will describe the diffuse radio sources found in and around the cluster within the context of the X-ray emission from the cluster core. In this paper \CORRS{we} assume a flat $\Lambda$CDM cosmology with $H_0 = 70$~km\,s$^{-1}$\,Mpc$^{-1}$, $\Omega_\mathrm{m} = 0.3$, and $\Omega_\Lambda = 1-\Omega_\mathrm{m}$. At the redshift of \spt, 1\arcmin~corresponds to 257~kpc.
~\\~\\
\section{Data}


\subsection{Radio}

\begin{table*}
    \centering
    \begin{threeparttable}
    \caption{\label{tab:obs} Details for the ASKAP and ATCA observations.}
    \begin{tabular}{c c c c c c c c c }
    \toprule
        Telescope & Dates & $N_\mathrm{ants} {}^\text{\tabft{\ref{tab:nants}}}$ & $B_\mathrm{max} {}^\text{\tabft{\ref{tab:maxbaseline}}}$ & $\tau {}^\text{\tabft{\ref{tab:time}}}$ & $\Delta\nu {}^\text{\tabft{\ref{tab:deltanu}}}$ & $\nu_\mathrm{c} {}^\text{\tabft{\ref{tab:nuc}}}$ & $\theta_\mathrm{max} {}^\text{\tabft{\ref{tab:maxangular}}}$ & $\sigma_\text{rms}$ \\
        {} & {} & {} & (m) & (min) & (GHz) & (GHz) & ($^\prime$) & ($\mu$Jy\,beam$^{-1}$) \\\midrule
        ASKAP & 18-07-2019 & 36 & 6\,000 & 600 & 0.288 & 0.943 & 49 & 60(31)[54] ${}^\text{\tabft{\ref{tab:noiseaskap}}}$ \\
        \multirow{2}{*}{ATCA(H75)} & \multirow{2}{*}{02-10-2009} & \multirow{2}{*}{$5 {}^\text{\tabft{\ref{tab:atcaants}}}$} & \multirow{2}{*}{89} & \multirow{2}{*}{42} & \multirow{2}{*}{2.049} & 5.5 & 6.0 & 103(100)[60]${}^\text{\tabft{\ref{tab:noiseatca}}}$\\
         & & & & & & 9.0 & 3.7 & 125 ${}^\text{\tabft{\ref{tab:noiseatca}}}$ \\ 
        \bottomrule
    \end{tabular}
    \begin{tablenotes}[flushleft]
    \footnotesize \item \emph{Notes.} 
    \item \ft{tab:nants} Number of antennas. \ft{tab:maxbaseline} Maximum baseline for observation. \ft{tab:time} Total integration time. \ft{tab:deltanu} Bandwidth. \ft{tab:nuc} Central frequency. \ft{tab:maxangular} Maximum angular scale that the observation is sensitive to.  \ft{tab:atcaants} The sixth antenna is not used due to poor $u$--$v$ coverage. \CORRS{\ft{tab:noiseaskap} Root-mean-square (rms) noise in the full-bandwidth uniform(robust $0.25$)[robust $+0.5$, tapered] ASKAP image. \ft{tab:noiseatca} rms noise in the uniform(robust $0.0$)[natural] ATCA images.}
    \end{tablenotes}
\end{threeparttable}
\end{table*}
\setcounter{ft}{0}

\subsubsection{Australian Square Kilometre Array Pathfinder}\label{sec:data:askap}

The ASKAP Evolutionary Map of the Universe \citep[EMU;][]{emu1} aims to observe the whole sky visible to ASKAP down to 10~$\mu$\jybeam, though observing and processing is still underway. ASKAP's phased array feeds \citep[PAF;][]{Chippendale2010,Hotan2014,McConnell2016} \CORRS{allow 36 independently-formed $\sim 1$~degree field-of-view (FoV) primary beams to be pointed on the sky \CORRStwo{within the $\sim 30$~deg$^2$ PAF FoV}}, making it a suitable instrument for survey work. Recently, a pilot set of observations for EMU were released, with calibrated visilibities made publicly available via the CSIRO \footnote{Commonwealth Scientific and Industrial Research Organisation} ASKAP Science Data Archive \citep[CASDA;][]{casda}. \CORRS{Prior to retrieving from CASDA, data go through radio frequency inteference (RFI) flagging, bandpass-calibration, and averaging} using the \texttt{ASKAPsoft} \footnote{\url{https://www.atnf.csiro.au/computing/software/askapsoft/sdp/docs/current/pipelines/introduction.html}} pipeline on the Pawsey Supercomputing Centre in Perth, Western Australia. For this observation (ID SB9351) ASKAP's 36 primary beams are formed into a ``6 by 6'' footprint, covering $\sim 6\degr \times 6\degr$. Each of these beams are processed and calibrated individually prior to co-adding in the image plane. PKS~B1934-638 is used for bandpass and absolute flux calibration, and the data are averaged to 1~MHz after calibration with a full bandwidth of 288~MHz. Observation details are presented in Table \ref{tab:obs}. After data are retrieved from CASDA, \CORRS{data are imaged using the widefield imager \texttt{WSClean} \footnote{\url{https://sourceforge.net/p/wsclean/wiki/Home/}} \citep{wsclean1,wsclean2}.} For each beam we generate deep images using the ``Briggs'' \citep{Briggs95} weighting with robustness parameter $+0.25$, splitting the full ASKAP bandwidth into six subbands of $\Delta\nu = 48$~MHz, all convolved to a common resolution of $16.2\arcsec \times 16.2\arcsec$ \CORRS{matching the resolution in the lowest-frequency band}. \CORRS{We use the multi-scale CLEAN algorithm to ensure extended sources are accurately modelled.} An additional full-bandwidth image is made for each beam, \CORRS{as well as an additional full-bandwidth image with uniform weighting resulting in a resolution of $8\arcsec \times 8\arcsec$.}\par
 
Initial imaging showed the presence of three discrete sources within the extended emission which prompted the need for compact-source--subtracted images. \CORRS{After trialling a number of methods---including source modelling via their spectral energy distributions (SEDs) and subtracting visibilities after imaging compact sources via a hard $u$--$v$ cut---we found that using CLEAN masks around the diffuse radio sources to be most effective.} For this, we re-image the calibrated data with robust $+0.25$ weighting, \CORRSthree{and using a CLEAN mask that excludes all diffuse emission we ensure that no diffuse emission is included in the CLEAN component model}. We then subtracted the \CORRS{frequency-dependent} CLEAN component model before imaging the data with a robust $+0.5$ weighting with an additional $20$~arcsec Gaussian taper to enhance the diffuse emission. We only generate four source-subtracted subband images to improve the signal-to-noise ratio ($\Delta\nu = 72$~MHz), specifically for the northern components of the emission. Note that cleaning prior to subtraction is done to the same depth as the robust $+0.25$ \CORRS{subbands}, and no residual sources are present in the subband images within the CLEAN mask region above the noise. \CORRS{We note for completeness that in the fullband images, residual emission can be seen around some (specifically extended) sources (see Fig. \ref{fig:radio}) though this full-band image is not used for flux density measurements}. All low-resolution subbands are convolved to a common resolution of $36\arcsec \times 36\arcsec$. \par
\CORRS{For each beam we perform a cross-match with a sky model generated from the GaLactic and Extragalactic All-sky MWA survey \citep[GLEAM;][]{wlb+15,gleamegc} and Sydney University Molonglo Sky Survey \citep[SUMSS;][]{Bock1999,Mauch2003} \CORRStwo{and extrapolate the measured SEDs of sources to 943~MHz. We find} a $\sim 5$~per cent standard deviation in measured flux densities compared to the extrapolated flux densities \CORRSthree{of} the $\sim 30$--40 brightest sources in each beam.}
\spt~falls within the half-power point of four of the 36 beams (6, 7, 13, and 14). We find the flux scale in the robust $+0.25$ images of beam 6 to be $\sim10$\% higher than the other three beams; \CORRS{as we are unable to identify the cause of this discrepancy and because its inclusion makes little difference to the resultant mosaic, beam 6 is removed} for all image sets. For each subband and the fullband image (robust $+0.25$ and source-subtracted), the three remaining beams are mosaicked, weighted by the square of the primary beam response (a circular Gaussian with full-width at half-maximum proportional to $1.09\lambda / D$, where $D$ is the dish diameter; A. Hotan, priv. comms.). Fig. \ref{fig:rgb} shows the fullband ASKAP data as contours, with the smaller highlight panels showing the source-subtracted data, with discrete sources labelled. A source-subtracted robust $+0.25$ image is presented in the left panel of Fig. \ref{fig:rgb} though is not used for quantitative analysis.

\begin{figure}[t]
    \centering
    \includegraphics[width=1\linewidth]{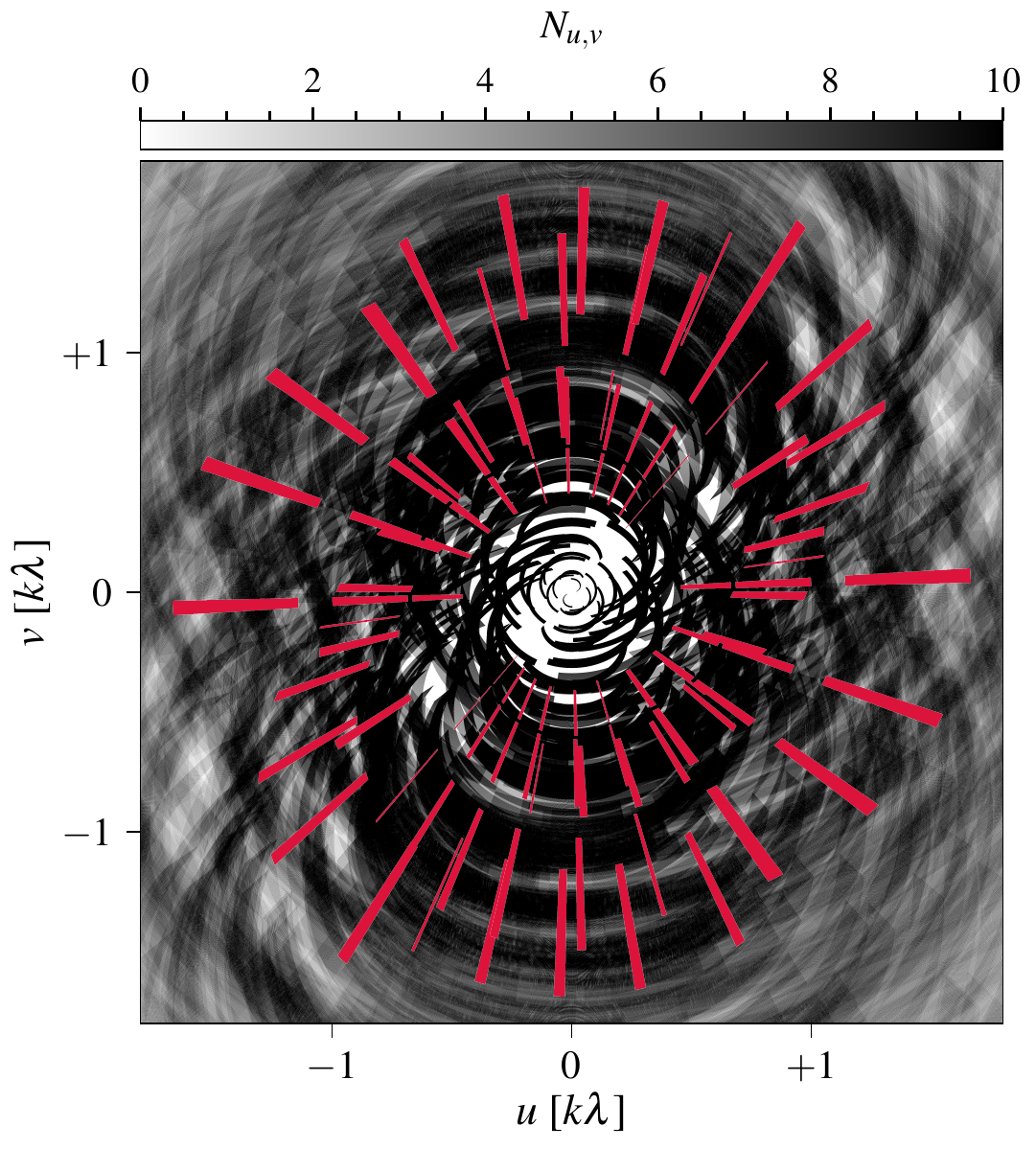}
    \caption{\label{fig:uv} \CORRSthree{ $u$--$v$ coverage for the 835.5~MHz ASKAP subband (black, $\Delta\nu=72$~MHz) and 5.5-GHz ATCA fullband (red, $\Delta\nu=2.049$~GHz) data within $-1.8~\text{k}\lambda \leq u \leq 1.8~\text{k}\lambda$ and $-1.8~\text{k}\lambda \leq v \leq 1.8~\text{k}\lambda$. The ASKAP data are represented as a density ($\lambda^{-2}$)}.}
\end{figure}

\subsubsection{Australia Telescope Compact Array 4~cm observations}\label{sec:data:atca}
\spt~was observed in 2009 with the ATCA using the Compact Array Broadband Back-end \citep[CABB;][]{cabb} for project C1563 (PI Walsh). The 4-cm band observations have dual instantaneous frequency measurements centered at 5.5 and 9~GHz with 2.049~GHz bandwidth. The cluster was observed for $\sim 42$~minutes in the H75 configuration (maximum baseline 4408~m; minimum baseline 31~m). The H75 configuration has a significant gap between the sixth antenna and the more compact core formed by the remaining antennas which results in a sampling function/point-spread function that can become difficult to accurately deconvolve when imaging \footnote{For antenna 6 to be useful in this configuration, additional observations in other ATCA configurations should be used.}. Additionally, because of the shorter observing time the baselines formed with the sixth antenna are significantly elongated in $u$--$v$ space. We remove antenna 6 prior to \CORRS{self-calibration and} imaging for this reason. The observation was performed in ``$u$--$v$ cuts'' mode to maximise $u$--$v$ coverage (see Fig. \ref{fig:uv} for a $u$--$v$ coverage plot with comparison to the ASKAP $u$--$v$ coverage). Observation details are presented in Table \ref{tab:obs}. Processing for these data utilise the \verb|miriad| software suite \citep{stw95} as well as \texttt{WSClean} and \texttt{CASA} for imaging and self-calibration. For initial bandpass and gain calibration, we follow the procedure of \citet{Duchesne2019} for the 4-cm data, using PKS~B1934$-$638 as the bandpass and absolute flux calibrator and PKS~1941$-$554 for phase calibration. \verb|WSClean| is used to generate model visibilities which are then used by the \verb|CASA| task \verb|gaincal| to perform phase-only self-calibration over three iterations, \CORRS{with solution intervals of 300, 120, and 60~s}. We only consider Stokes \textit{I} for this analysis as the resolution of the H75 configuration without antenna 6 is too poor for resolved polarimetry and the observation length prohibits good parallactic angle coverage. \par
\begin{figure*}[t]
    \centering
    \begin{subfigure}{0.5\linewidth}
    \includegraphics[width=1\linewidth]{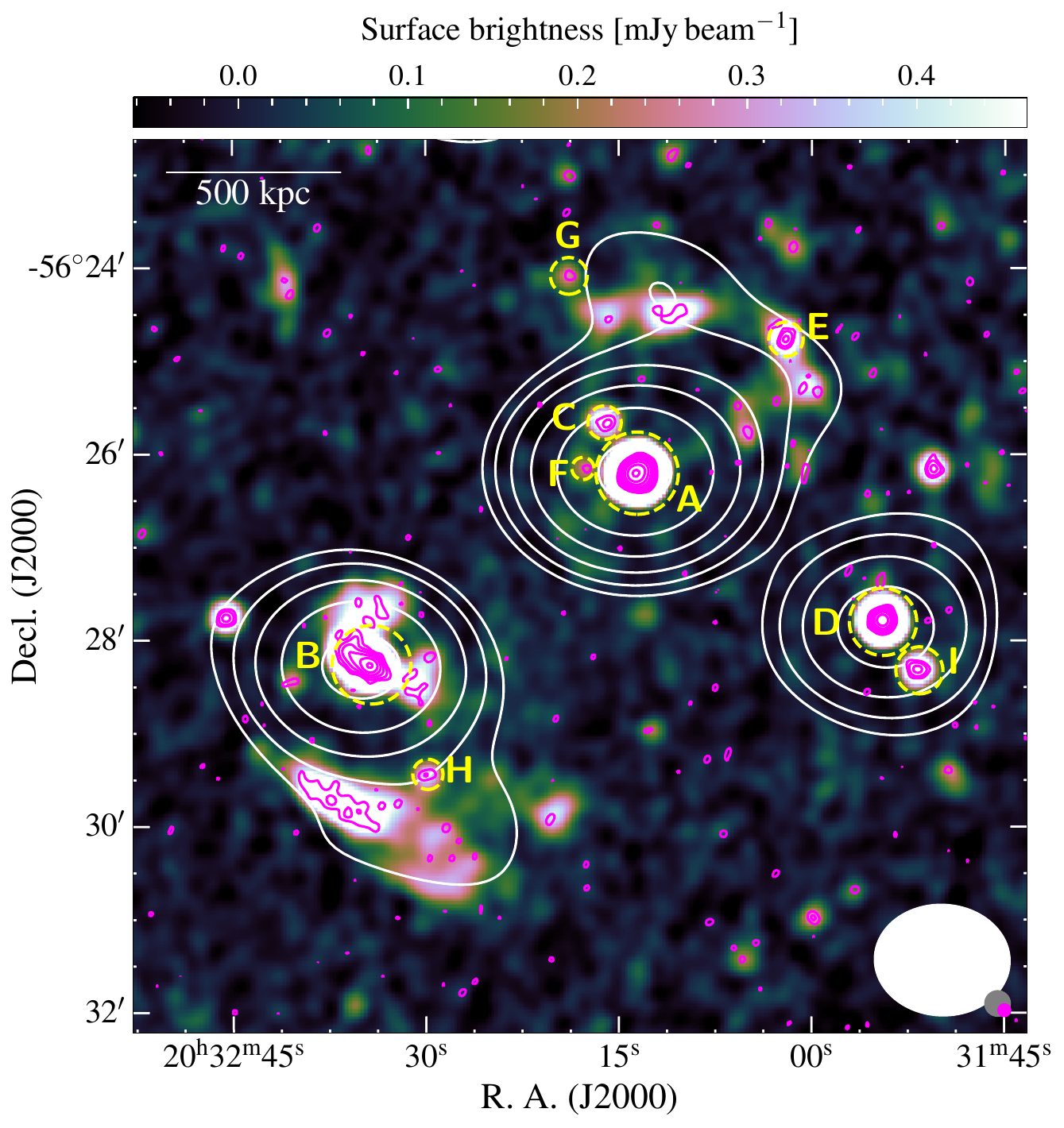}
    \caption{\label{fig:radio:fullres}}
    \end{subfigure}%
    \begin{subfigure}{0.5\linewidth}
    \includegraphics[width=1\linewidth]{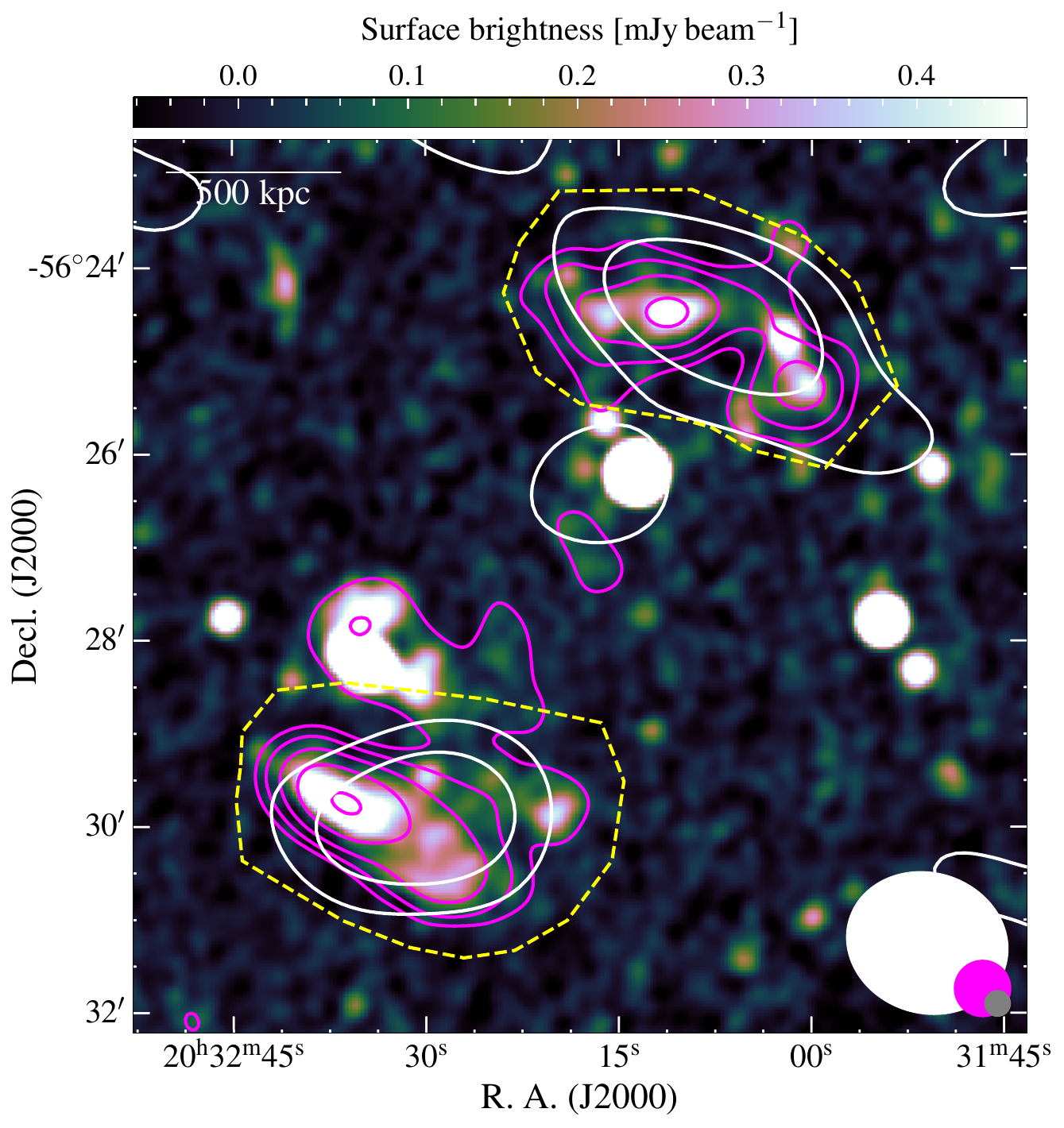}
    \caption{\label{fig:radio:lowres}}
    \end{subfigure}%
    \caption{\label{fig:radio} \CORRS{Radio data for \spt. The background image is the  fullband ASKAP $16.2\arcsec$ data. All contours start at $3\sigma_\mathrm{rms}$ and increase with factors of 2. The contours are as follows: \subref{fig:radio:fullres} ASKAP $8\arcsec$, magenta, starting at $180$~$\mu$Jy\,beam$^{-1}$; ATCA 5.5-GHz robust 0.0 image, white, starting at $300$~$\mu$Jy\,beam$^{-1}$. \subref{fig:radio:lowres} ASKAP $36\arcsec$ source-subtracted, magenta, starting at $162$~$\mu$Jy\,beam$^{-1}$; ATCA 5.5-GHz source-subtracted naturally-weighted, white, starting at $180$~$\mu$Jy\,beam$^{-1}$. Yellow labels in \subref{fig:radio:fullres} are discrete sources discussed in the text (Section \ref{sec:results:others}). Yellow, dashed regions in \subref{fig:radio:lowres} are the flux density integration regions for the NW and SE relics. The beam shapes are shown as ellipses in the bottom right corner, with grey corresponding to the background map. The linear scale in the top left corner is at $z=0.284$.}}
\end{figure*}

While radio relics are rarely detected above 4~GHz (except in the particularly bright cases e.g. \citealt{srm+01,Loi2017}), initial imaging \CORRS{at a `Briggs' robust 0.0 weighting (Fig. \ref{fig:radio:fullres})} revealed residual emission in the vicinity of the relic sources. \CORRStwo{To determine the nature of this emission, we subtracted the compact components. This is done by imaging the data with a uniform weighting, creating a CLEAN mask that targets only the brightest sources (A, B, and D in Fig.~\ref{fig:radio:fullres}), and CLEANing down to $1\sigma_\mathrm{rms}$ within this mask. The CLEAN model components of these bright, compact sources are subtracted, revealing residual diffuse emission. The remaining discrete sources shown in Fig.~\ref{fig:radio:fullres} are not subtracted at this stage. This process results in some over-subtraction of the diffuse emission at the location of source B, though this over-subtraction will be minimal due to the lack of evidence of the diffuse emission in the uniform image. Within the full-width at half maximum (FWHM) of the primary beam no additional residuals are seen in the source-subtracted image to suggest further contamination from poor subtraction except at the location of source F (see Fig.~\ref{fig:radio:lowres}).}  \par
To maximise sensitivity here we image the full 2-GHz band with a \CORRS{natural weighting}. The resultant source-subtracted 5.5-GHz image is shown in Fig. \ref{fig:radio:lowres} overlaid on the ASKAP data. Note that 5.5-GHz emission is seen from source F (denoted in Fig. \ref{fig:rgb}) and some low-level residual emission remains from other cluster sources. \CORRS{Additionally, the subtraction of source B is not perfect.} Additional \CORRS{uniformly weighted}, non--source-subtracted images were made at 5.5- and 9-GHz to measure the SEDs of sources A and B.

\subsection{X-ray}\label{sec:data:xmm}

\spt~was observed by \xmm\ using the European Photon Imaging Camera (EPIC; \citealt{turner2001} and \citealt{struder2001}) for $32$~ks (Obs. ID 0674490401). The data were \CORRS{retrieved from the archive and} reprocessed using the Science Analysis System version 18.0.0, using the latest calibration files available as of December 2019. High energy particle contamination was reduced by removing events for which the \verb?PATTERN? keyword was $>4$ and $>13$ for the MOS1,2 and PN cameras, respectively. Observation intervals affected by flares were identified and removed from the analysis following the prescriptions detailed in \cite{pratt2007}. The cleaned exposure times are $27.7$~ks and $20.7$~ks for MOS1,2 and PN cameras, respectively. Point sources were identified following the procedures described in \citet{bogdan13}, and removed from the subsequent analysis.

After these procedures, data taken by the three cameras were combined to maximise the signal-to-noise, and arranged in data-cubes. Exposure maps and models of the sky and instrumental background were computed as described in \citet{bourdin08}, \citet{bourdin13} and \citet{bogdan13}. %

\section{Results}

\begin{table*}
    \centering
    \begin{threeparttable}
    \caption{\label{tab:interlopers} Compact/intervening radio sources in the field of \spt.}
    \begin{tabular}{c c c c c}\toprule
        ID. & Name & $z$ & $\alpha$ \tabft{\ref{tab:alpha}} & $q$ \tabft{\ref{tab:q}} \\\midrule
        A   & 2MASS~J20321413$-$5626117 & $0.2844 \pm 0.0002$ \tabft{\ref{tab:redshift:A}} & $-0.64\pm0.07$& $-0.07\pm0.03$ \\
        B   & 2MASX~J20323421$-$5628162 & - & $-0.38\pm0.08$& $-0.12\pm0.04$ \\
        C   & 2MASS~J20321605$-$5625390 & $0.0621 \pm 0.0001$ \tabft{\ref{tab:redshift:A}} & $-1.24\pm0.81$& - \\
        D   & SUMSS J203154$-$562749 & - & $-0.92\pm0.05$& - \\
        E   & WISEA J203202.07$-$562445.0 & - & $-1.07\pm0.59$& - \\
        F \tabft{\ref{tab:f}}   & 2MASS J20321740$-$5626084 & $0.2841 \pm 0.0002$ \tabft{\ref{tab:redshift:A}} & - & - \\
        G   & WISEA J203219.10$-$562405.9 & -  & $-0.06\pm1.04$& - \\
        H   & WISEA J203229.85$-$562925.9 & - & $-1.55\pm0.98$& - \\
        I   & WISEA J203151.76-562818.3 & - & $-0.29\pm0.57$& - \\
        \bottomrule
    \end{tabular}
    \begin{tablenotes}[flushleft]
    \footnotesize \item \emph{Notes.} \ft{tab:f} No SED measured as it is too faint in the ASKAP subbands. \ft{tab:alpha} $S_\nu \propto \nu^{\alpha}$. \ft{tab:q} $S_\nu \propto \nu^{\alpha}e^{q\ln \nu^2}$. \ft{tab:redshift:A} \citet{Ruel2014}.
    \end{tablenotes}
\end{threeparttable}
\end{table*}
\setcounter{ft}{0}

\begin{figure}[!t]
    \centering
    \includegraphics[width=1\linewidth]{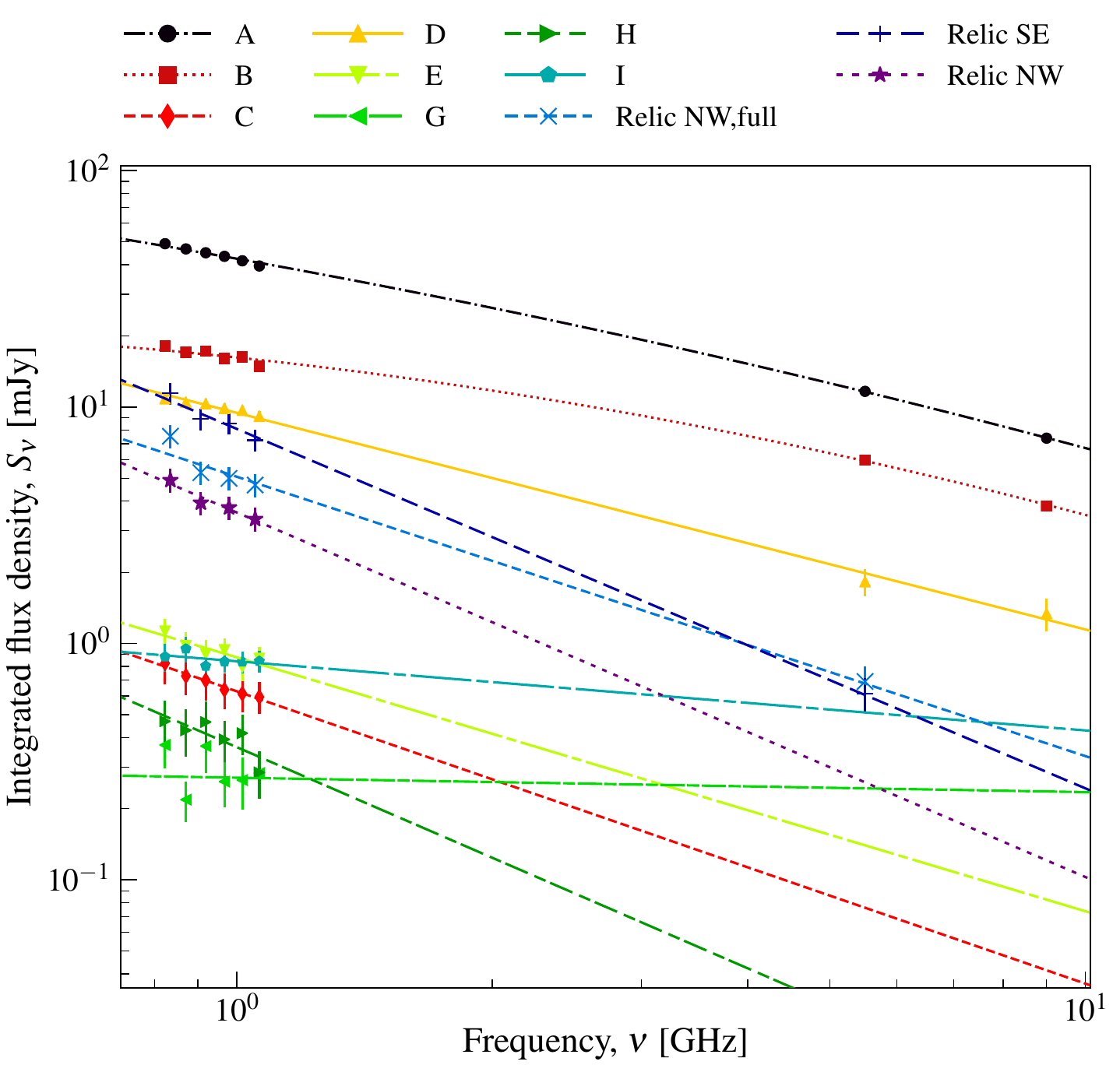}
    \caption{\label{fig:seds} SEDs of compact radio sources in the cluster field. Sources A and B are fitted with generic curved power law models while the remaining sources are with a normal power law. \CORRS{Model parameters are reported in Table \ref{tab:interlopers} and \ref{table:relics} for the discrete sources and relics, respectively.}}
\end{figure}

\begin{table}[!t]
    
    \centering
    \caption{\label{table:relics} Relic emission flux density measurements and physical properties.}
    \resizebox{\linewidth}{!}{\begin{tabular}{c c c c c c }\toprule
Relic & $\nu$ & $S_\nu$ & $\alpha$ & LLS & $\sigma_\text{rms} {}^{\text{\tabft{\ref{tab:rms}}}}$ \\
          & (MHz) & (mJy) & & (kpc) & ($\mu$Jy\,beam$^{-1}$) \\\midrule
\multirow{5}{*}{SE} & 835.5 & $11.4 \pm 0.7$ & \multirow{5}{*}{\alphaSonly} & \multirow{5}{*}{731} & 109 \\
 & 907.5 & $8.92 \pm 0.53$ & & & 84 \\
 & 979.5 & $8.54 \pm 0.51$ & & & 75\\
 & 1051.5 & $ 7.24 \pm 0.44$ & & & 68 \\
  & 5500.0 & $0.61 \pm 0.10$ & & & 60 \\\midrule

\multirow{5}{*}{NW} & 835.5 & $4.91 \pm 0.37$ & \multirow{5}{*}{\alphaNEonly} & \multirow{5}{*}{447} & 100 \\
 & 907.5 & $3.95 \pm 0.30$ & & & 82 \\
 & 979.5 & $3.76 \pm 0.28$ & & & 77 \\
 & 1051.5 & $ 3.35 \pm 0.25$ & & & 72 \\
 & 5500.0 & $ < 0.69 \pm 0.10$ & & & 60 \\\midrule
 
\multirow{5}{*}{NW, full} & 835.5 & $ 7.52 \pm 0.53 $ & \multirow{5}{*}{\alphaNonly} & \multirow{5}{*}{860} & 101 \\
 & 907.5 & $5.28 \pm 0.38$ & & & 82 \\
 & 979.5 & $4.99 \pm 0.36$ & & & 77 \\
 & 1051.5 & $ 4.68 \pm 0.36$ & & & 72 \\
  & 5500.0 & $ 0.69 \pm 0.10 $ & & & 60 \\\bottomrule
    \end{tabular}}\\[0.25em]
    {\footnotesize \textit{Notes.} \ft{tab:rms} Average rms over the source.}
\end{table}
\setcounter{ft}{0}

The two relics are clearly detected in the ASKAP data as well as the source-subtracted 5.5-GHz ATCA data. Fig. \ref{fig:rgb} shows the location of the relic sources (top right, bottom right panels), and the top panel in that figure indicates where additional emission from the NW relic may be located (surrounding source E). The two relics are located $\sim850$ and $\sim800$~kpc from the reported cluster centre in the SE and NW, respectively. Fig. \ref{fig:radio} shows the radio data used here, showing the full resolution ASKAP data with lower resolution ASKAP contours overlaid along with the source-subtracted 5.5-GHz ATCA data. From the ASKAP data we estimate the projected extent of the emission: SE relic, 2.77\arcmin~(731~kpc at $z=0.284$), NW relic, 1.71\arcmin~(447~kpc, main eastern portion) and 1.58\arcmin~(413~kpc, secondary western portion). While there is a compact source (E) within the emission to the west of the NW relic, it is not clear whether the emission surrounding E is associated with the NW relic or not. In the event that it is, \CORRS{then the relic forms a continuous structure with a} total projected linear size is 860~kpc. 
\subsection{Other radio sources}\label{sec:results:others}
The ASKAP data reveal a number of radio sources projected onto the cluster system. As well as the diffuse sources, we will discuss an additional \CORRS{nine} sources within the images, labelled \CORRS{A--I},  shown on Fig. \ref{fig:rgb} and Fig. \ref{fig:radio}. We note the source names and redshifts (where available) in Table \ref{tab:interlopers}. \CORRS{For sources A, B, and D we measure the integrated flux densities in the ASKAP and ATCA 5.5- and 9-GHz data}. The spectra of \CORRS{A and B} show curvature and are fit with a generic curved power law model \CORRS{of the form \begin{equation}
    S_\nu \propto \nu^\alpha e^{q \left(\ln \nu\right)^2} \, ,
\end{equation}
where $q$ is a measure of the curvature and for $q=0$ the model reduces to a standard power law ($S \propto \nu^\alpha$). For the remaining sources we measure flux densities only from the ASKAP data and all sources except A and B are fit with generic power law models.} 

Fig. \ref{fig:seds} shows SEDs of each of these sources along with the fitted models, \CORRS{and the fitted model parameters are shown in Table \ref{tab:interlopers}}. \CORRS{As sources A, B, and \CORRSthree{D$+$I are point-like} sources in the ATCA uniformly weighted images, we use the \texttt{aegean} source-finding software \footnote{\url{https://github.com/PaulHancock/Aegean/tree/master/AegeanTools}} \citep{hmg+12,hth18} to obtain flux density measurements at 5.5 and 9 GHz.} \CORRSthree{As source D and I become a single point-like source in the ATCA data, we subtract the extrapolated contribution of source I from the \texttt{aegean} measurement to obtain flux density measurements at 5.5 and 9~GHz for source D alone. Note that the contributions of sources C and F to the measurement of source A in the ATCA data are significantly less than the errors and are not considered further.} \CORRStwo{Due to the complexity of source B and \texttt{aegean} failing to fit the fainter discrete sources, in} the ASKAP images we directly measure \CORRS{the} flux densities by integrating over regions containing the relevant sources. \par

\subsection{Radio relic spectral energy distribution}\label{sec:results:sed}
Within the ASKAP source-subtracted, tapered subband images ($\Delta\nu = 72$~MHz) we integrate the flux density within bespoke polygon regions for each relic source \CORRStwo{(shown as yellow dashed regions in Fig.~\ref{fig:radio:lowres})}. For estimating the uncertainty, we use the Background And Noise Estimation tool \citep[\texttt{BANE};][]{hmg+12} to estimate the map rms noise and add an additional uncertainty based on the flux scale and primary beam correction of 5\%. Due to the complex nature of the NW relic source, we measure two different regions, the first being only the most eastern portion of the NW relic, and the second including the emission surrounding source E. \par
Additionally, we use the residual, source-subtracted 5.5-GHz maps as described in Section \ref{sec:data:atca} to measure the integrated flux densities of the two relics in the 5.5-GHz map, also integrating the flux within bespoke polygon regions. The regions used here are larger than those used for the ASKAP data due to the difference in resolution. For the first NW relic measurement (i.e. the eastern component) we consider the full 5.5-GHz measurement as an upper limit as we are unable to subtract the contribution from the emission surrounding source E.  We add an additional flux scale uncertainty of 10\% for these data due to the inherent uncertainty in the source-subtraction with incomplete $u$--$v$ data \CORRS{combined with the low resolution of the H75 array and ratio of the primary beam and restoring beams ($\sim 5:1$) and considering the relic sources lie toward the FWHM of the primary beam}. Additionally, we subtract the estimated flux density contribution from source E, G, and H, which are not subtracted in visibilities for the 5.5-GHz data. \par

Fitting a normal power law to the two relic SEDs yields spectral indices of \alphaS, \alphaNE, and \alphaN, where $\alpha_\mathrm{NW,full}$ is the spectral index for the NW relic including the emission surrounding source E. We estimate the 1.4-GHz power of the relics from these spectral indices: $P_{1.4}^{\text{SE}} = (\powerS)\times 10^{23}$~W\,Hz$^{-1}$, $P_{1.4}^{\text{NW}} = (\powerW)\times 10^{23}$~W\,Hz$^{-1}$, and $P_{1.4}^{\text{NW,full}} = (\powerN)\times 10^{23}$~W\,Hz$^{-1}$. The measured data and fits are shown in Fig. \ref{fig:seds} and presented in Table \ref{table:relics} for the relic sources.

\subsection{\CORRS{X-ray cartography}}

\begin{figure*}[!t]
    \begin{center}
        \begin{subfigure}[b]{0.5\linewidth}
        \includegraphics[width=1\linewidth]{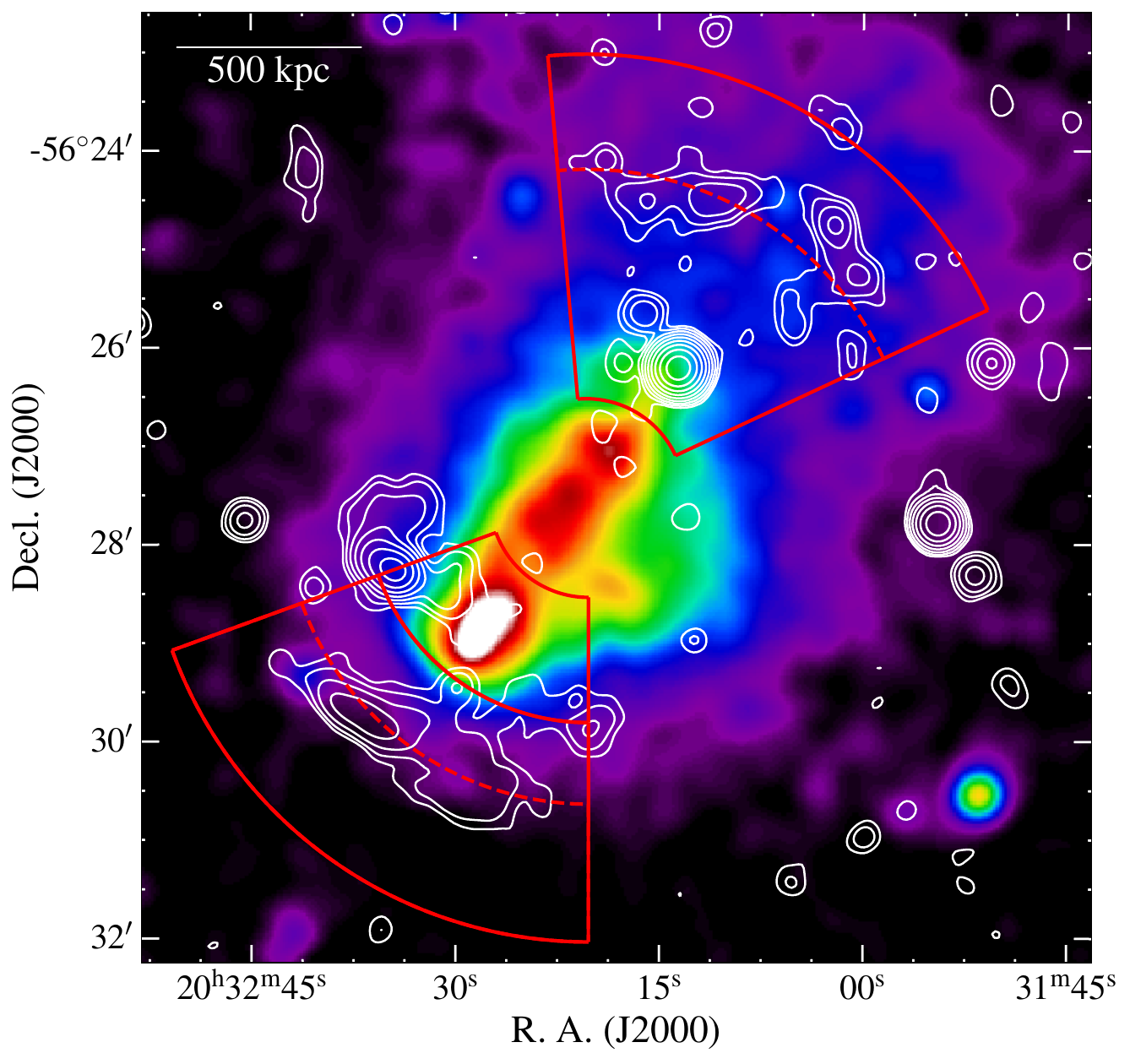}
        \caption{\label{fig:xmm_wt_kt:wt}}
        \end{subfigure}%
        \begin{subfigure}[b]{0.5\linewidth}
        \includegraphics[width=1\linewidth]{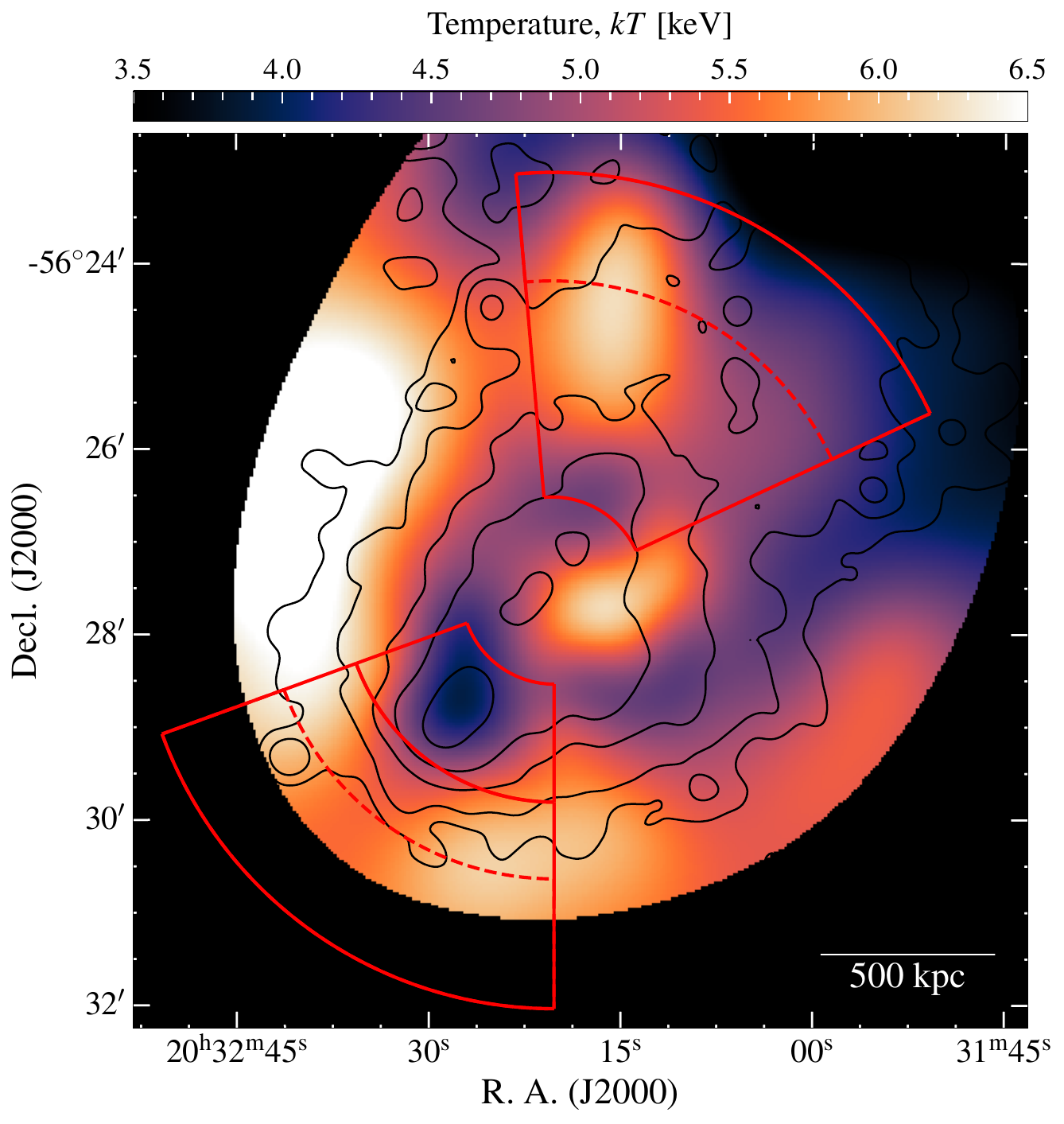}
        \caption{\label{fig:xmm_wt_kt:kt}}
        \end{subfigure}%
    \end{center}
    \caption{\label{fig:xmm_wt_kt} \subref{fig:xmm_wt_kt:wt} exposure-corrected, background-subtracted, and smoothed X-ray image of \spt. \CORRS{The white contours are the $16.2^{\prime\prime}$ fullband ASKAP image, starting at $3\sigma_\text{rms}$ ($\sigma_\text{rms} = 31$~$\mu$Jy\,beam$^{-1}$)}. The sectors shown in red are those used for surface brightness profiles. The linear scale in the bottom right corner is at $z=0.284$. \subref{fig:xmm_wt_kt:kt} Temperature map of the cluster, with sectors used for surface brightness profile analysis overlaid as in \subref{fig:xmm_wt_kt:wt}. The dashed red arcs within the sectors are at the peak radio locations of the relics, and the solid red arc in the SE sector is the location of a cold front. The black contours are from the smoothed X-ray image. \CORRS{The wavelet filtering algorithms used to obtain these maps are described in \citealt{bourdin2004,bourdin08,bourdin13}.}}
\end{figure*}

The wavelet-filtered, exposure-corrected, and smoothed X-ray image, overlaid with contours from the ASKAP data, is shown in Fig. \ref{fig:xmm_wt_kt:wt}. The X-ray morphology of the cluster appears disturbed, with a major axis of elongation along the SE--NW direction. There is a clear bright structure in the centre of the cluster with a tail elongated in the NW direction. 

\CORRS{We investigated the projected temperature distribution of the ICM by producing a temperature map using the wavelet filtering approach described in \citet{bourdin2004} and \citet{bourdin08}; the result is shown in Fig. \ref{fig:xmm_wt_kt:kt}.}

\CORRS{The NW region does not show any particular features, the mean temperature being $\sim 5$ keV with no strong spatial variation. }
\CORRS{The SE region, in contrast, is characterised by the presence of a distinct cold spot at $\sim 4$ keV, which is spatially coincident with the X-ray brightness peak. This cold spot is surrounded in the SW direction by a bow-like hotter region at $\sim 6$ keV. The transition between the cold and the hot regions corresponds to a clear drop in the surface brightness visible in the cluster image. This configuration suggests the possible presence of a cold front, resembling that found by \citet{bourdin08} in Abell 2065. However, the radio relic contours follow the SE bow-like hot region, and could therefore signal the presence of a shock related to the merger.}

\CORRS{The typical signature of a cold front is a sudden and sharp drop in the ICM brightness map that produces a change of slope in the surface brightness radial profile. This change of slope is translated into a clear jump-like feature in the deprojected density profile, across which the pressure remains constant i.e. the pressure profile does not show features across the cold front. In contrast, the presence of a shock produces a similar signature in the surface brightness and density profiles, but also a jump in the pressure profile. To clarify the nature of the bow-like structure we extracted surface brightness and temperature profiles from annular sectors shown in both panels of Fig. \ref{fig:xmm_wt_kt} using the radio contours as anchors.}

\subsection{X-ray and radio surface brightness profiles}

\begin{figure}[!t]
    \centering
    \begin{subfigure}{1\linewidth}
    \includegraphics[width=1\linewidth]{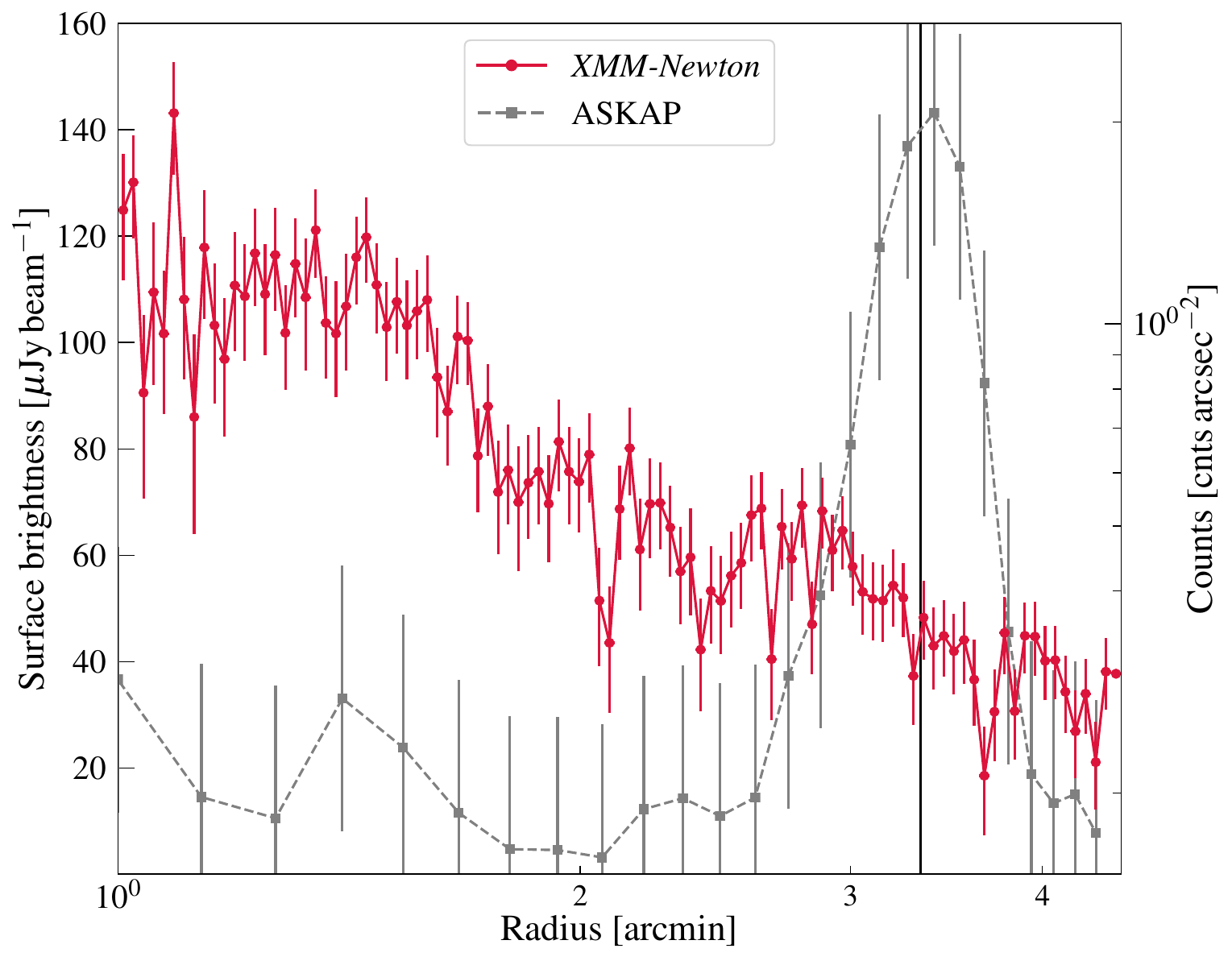}
    \caption{\label{fig:profile:nw} NW profile.}
    \end{subfigure}\\%
    \begin{subfigure}{1\linewidth}
    \includegraphics[width=1\linewidth]{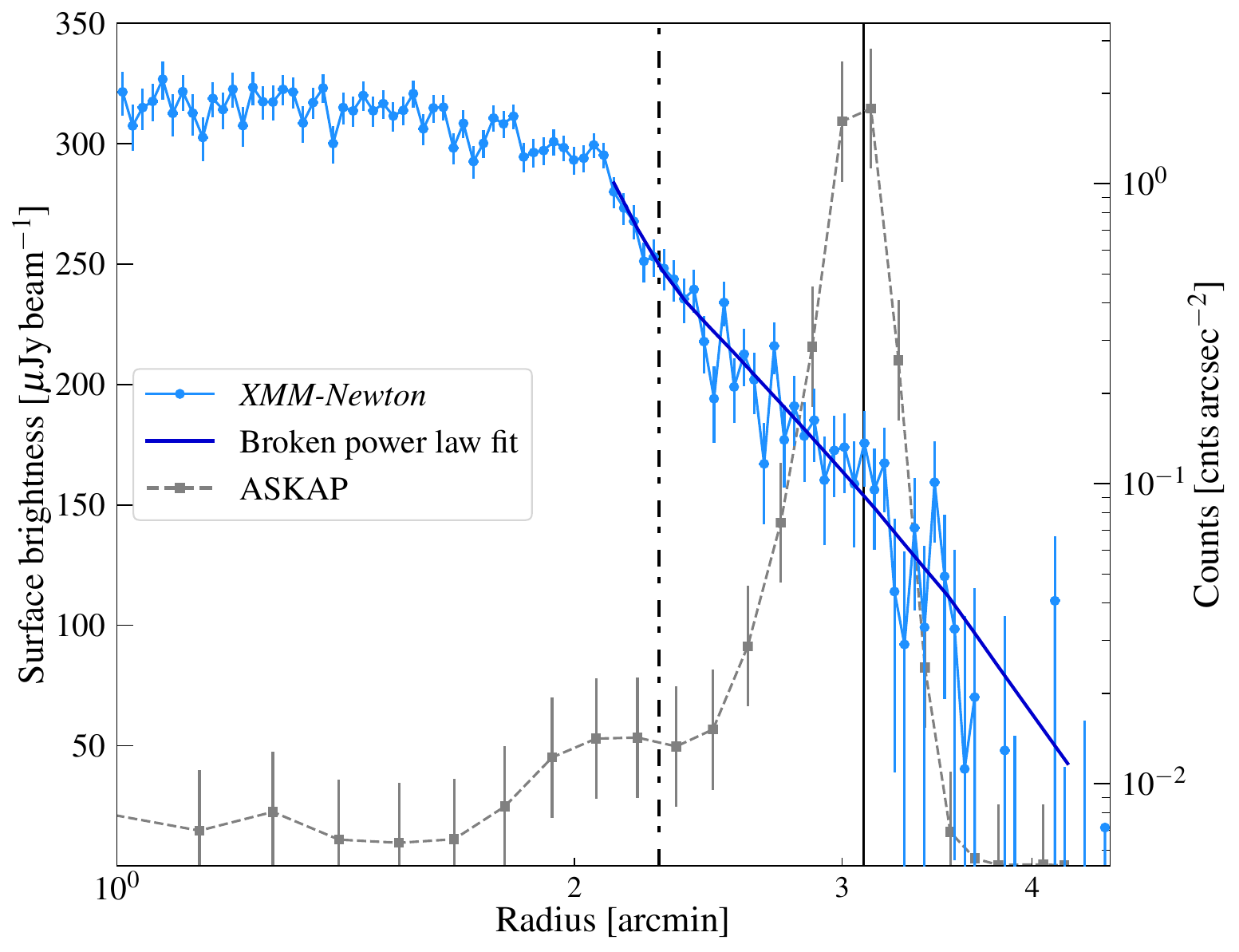}
    \caption{\label{fig:profile:se} SE profile.}
    \end{subfigure}%
    \caption{\label{fig:profile} X-ray and radio surface brightness profiles in the directions of the NW \subref{fig:profile:nw} and SE \subref{fig:profile:se} relic regions. \CORRS{The solid, vertical lines are the peak locations of the radio emission, and the dot-dashed, vertical line in \subref{fig:profile:se} is the location of a cold front. The broken power law fit to the SE profile is shown in \subref{fig:profile:se} as a solid, dark blue line.}}
\end{figure}

\begin{figure}
    \centering
    \includegraphics[width=1\linewidth]{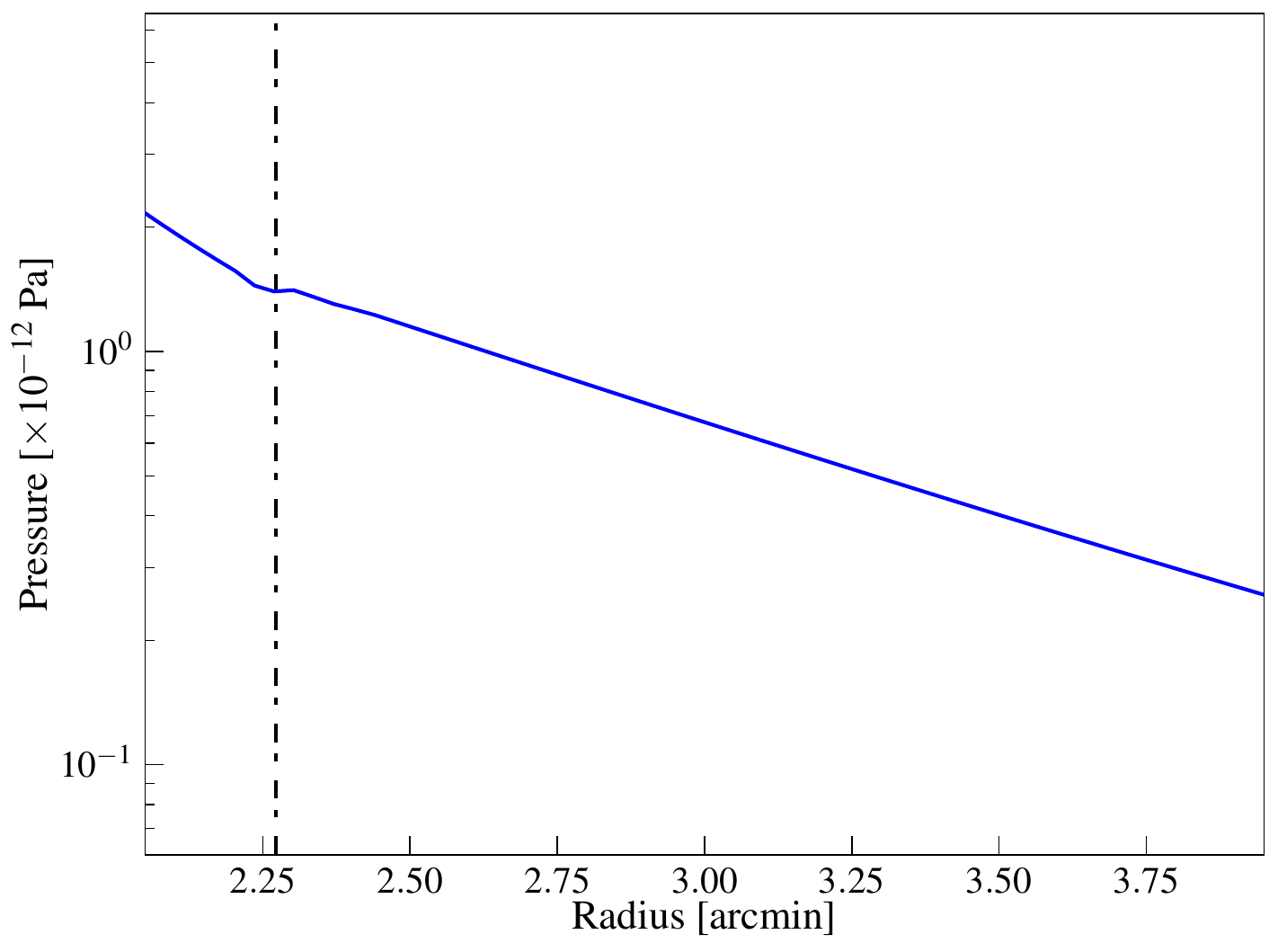}
    \caption{\label{fig:pressure} Derived pressure profile through the SE sector as discussed in the text. The dot-dashed, vertical line is the cold front location as in Fig. \ref{fig:profile:se}. Note the radial range covers 2--4~arcmin.}
\end{figure}

\CORRS{We extracted background-subtracted, exposure- and vignetting-corrected X-ray brightness profiles \CORRS{in the $[0.5-2.5]$ keV band} along the sectors highlighted in Fig. \ref{fig:xmm_wt_kt}. \CORRS{The profiles were binned using a logarithmic bin factor of 1.015, with a mean bin width of 2~arcesc corresponding to 9.2~kpc}. These profiles are shown in Fig. \ref{fig:profile}, where we also overlay the radio surface brightness profiles in the same regions from the full bandwidth ASKAP image as shown in Fig. \ref{fig:radio}. The presence of a sudden change of the slope in the X-ray surface brightness profile could potentially reveal the existence of a shock.}
\CORRS{Interior to the position of the radio relic, the X-ray surface brightness profile of the NW sector is noisy and does not show significant features. The radio relic profile peaks at $\sim 3.3$~arcmin, where we again do not see any correspondence with any obvious feature in the X-ray profile.}

\CORRS{The SE profile appears flat below 2 arcmin and then starts dropping. The point where the SE profile slope changes  corresponds to the end of the very bright spot that can be seen in Fig. \ref{fig:xmm_wt_kt:kt}. There is a first change of surface brightness slope at $\sim 2.5$~arcmin matching the position of the bow-like structure.  We identified the exact position by fitting the data with a broken power law model in the $[2-4]$ range, following the prescription of \citet{bourdin08}. The EPIC \CORRSthree{point-spread function (PSF)} is taken into account in the fitting procedures. The result is shown with a solid blue line in Fig. \ref{fig:profile:se}. The change of the slope occurs at \CORRS{$2.272 \pm 0.014$~arcmin ($578 \pm 4$~kpc)}. This position is shown with a red solid line in the SE sector in the left panel of Fig. \ref{fig:xmm_wt_kt}, and corresponds to the sharp edge in the brightness map and to the bow-like structure in the temperature map.}
\CORRS{We extracted the temperature in the two bins of the SE annular sector shown in Fig. \ref{fig:xmm_wt_kt:kt} obtaining $4.08_{-0.36}^{+0.38}$ keV and $8.50_{-3.51}^{+9.47}$ keV in the upper and lower bin, respectively. We then deprojected the density and the temperature profiles following \citet{bourdin08}. We investigated the nature of the slope change by looking at the pressure profile, shown in Fig. \ref{fig:pressure}. The profile across this feature appears to be smooth, as we would expect in the case of a cold front.}
\CORRS{We also fitted the surface brightness profile using the density model proposed by \citet[][their equation 3]{vikh06} and set the core term, $n_{02}$, to 0. We compared the result of this fit with the broken power law by using the F-test which yields a value of $16.36$ and probability of $0.0049$. These values suggest that the use of a double power law model significantly improves the fit. For the reasons discussed above we can interpret this first change in surface brightness slope as being due to the presence of a cold front.}

\CORRS{The radio profile in the SE peaks at $\sim 3.3$~arcmin. The X-ray surface brightness profile slope seems to change again around this point and the model of the broken power law is on average above the data points beyond $3.3$ arcmin. Unfortunately, the data are insufficiently deep to determine the position of this feature, or to extract a temperature profile. The radio relic peak matching a possible change of slope of the X-ray brightness suggests the presence of a shock, and a deeper X-ray observations of the cluster could allow us to confirm this.}

Interestingly, the SE profile and the overall scenario suggested by the SE profile are very similar to the case of Abell 2146 \citep{russell1,russell2}. The geometry of the cluster is similarly disturbed and our SE profile resembles the profiles shown in Fig. 5 of \citet{russell1}. Additionally, \citet{HlavacekLarrondo2018} report large-scale radio emission in Abell~2146, though it is not clear whether the radio emission found corresponds to radio relics, a radio halo, or a combination thereof \citep[see also][]{Hoang2019}.

\section{Discussion}

\subsection{Connection of relic emission to active cluster radio sources}\label{sec:discussion:connection}
To date a number of radio relic-like sources have been found in clusters with emission connected to apparently active radio galaxies \citep[see e.g.][]{Bonafede2014,Shimwell2015,vanWeeren2017,gasperin_gentle_2017}, however, the exact mechanism that energises the particle population in these cases is not clear. We will briefly consider the possibility that the SE and NW relic sources in \spt~are associated with active cluster sources, where a shock passing through a population of low-energy electrons (e.g. old radio lobes) imparts energy to re-accelerate the electron population (either through DSA or adiabatic compression).

\textit{SE relic.} Source B is an FR-I radio galaxy associated with 2MASX~J20323421$-$5628162 with jets oriented roughly NE--SW (the same orientation as the SE relic source, \CORRS{see Fig. \ref{fig:radio:fullres}}). The host galaxy has no reported redshift, but has a lower magnitude than the surrounding galaxies in \spt. The host galaxy has \textit{WISE} \footnote{Wide-field Infrared Survey Explorer; \citet{wright_wide-field_2010}} $W1$ ($W2$) magnitudes of $13.597\pm0.025$ ($13.365\pm0.030$) compared to the mean for the 31 cluster members of $15.8\pm1.0$ ($15.6\pm1.0$). We do not consider that this galaxy should be significantly brighter than the cluster members, and given it is not near the centre of the cluster it is unlikely this is the BCG of \spt. Given the difference in \textit{WISE} magnitudes, we \CORRS{suspect} that this galaxy is not a cluster member. Similarly, source C ($z=0.0621$; \citealt{Ruel2014}) has \textit{WISE} magnitudes $14.057\pm0.037$ ($13.894\pm0.044$), at the redshift of the reported foreground cluster Abell~3685 ($z=0.0620$; \citealt{sr99}), though note that Abell~3685 is considered a richness [0] cluster, but only has a single galaxy with a confirmed redshift. \CORRS{We conclude that there is no cluster at the reported redshift and location of Abell~3685. If the SE relic originated in source B, the lack of physical connection would suggest some episodic activity wherein the the SE relic would have faded and would require some ICM-related re-acceleration from shock physics (i.e. a relic).}

\textit{NW relic.} Source E is the likeliest candidate for a host for the NW relic, however, the orientation of the east component of the NW relic is \CORRS{almost} perpendicular with respect to the western component \CORRS{(see e.g. Fig. \ref{fig:radio:fullres})}, which would require E be a wide-angle tailed (WAT) radio source. Such WAT sources typically occur in cluster environments \citep[e.g.][]{Mao2009,Pratley2013} and it is thought the ram pressure of the ICM on the lobes generates the observed morphology \citep[][]{Burns1998}. Usually, this occurs during infall into the cluster \CORRS{but such a morphology may also occur after a radio galaxy has passed through the cluster centre. The asymmetry in the two NW relic components with respect to source E suggests that this is less likely to be an active WAT source, and the spectral index of the western relic component is sufficiently steep to preclude its classification as a normal radio lobe.}

\subsection{A merging system}\label{sec:discussion:merging}

{The X-ray morphology clearly shows that this is a disturbed system. The X-ray image confirms the merging axis along the SE--NW direction suggested by the position of the radio relics.}
 Assuming DSA and using the derived spectral indices for the two relic sources, we can estimate the Mach number, $\mathcal{M}$, of any shock that has generated them via \citep[see e.g.][]{Blandford1987} \begin{equation}
    \mathcal{M} = \sqrt{\dfrac{2\alpha_\mathrm{inj} + 3}{2\alpha_\mathrm{inj} -1}} \, ,
\end{equation}
where $\alpha_\mathrm{inj}$ is the emission injection index of the power law distribution of relativistic electrons. Without making an assumption on the injection index, we use the integrated spectral indices of the relics to determine lower limits to the Mach numbers, finding \CORRS{$\mathcal{M}_\mathrm{SE} \geq 1.7\pm0.1$, $\mathcal{M}_\mathrm{NW,full} \geq 2.0 \pm 0.1$, and  $\mathcal{M}_\mathrm{NW} \geq 1.7\pm0.6$}. For the full NW source and the SE source, the integrated spectral indices are consistent with those found for other double radio relics \footnote{Double radio relics in this context are the same as the ``double radio shocks'' (dRS) described in \citet{vda+19}.} (hereafter dRS) sources (with a mean value of $\langle \alpha \rangle \approx -1.2$ for the current sample of dRS sources; see Appendix \ref{sec:appendix:data}), and by construction the derived Mach numbers are as well \citep[see e.g. figure 24 of][]{vda+19}.

{Unfortunately the X-ray data is insufficient to probe for evidence of shocks. While some relic and double relic sources have been found to correlate to an X-ray--detected shock \citep[e.g.][]{Mazzotta2011,Finoguenov2010,Akamatsu2012,Akamatsu2013,botteon_mathcal_2016,botteon_shock_2016}, this is not true for all relic sources. This could partly be due to data quality issues, as often the exposures are insufficiently deep to allow extraction of temperature profiles at the large cluster-centric distances at which radio relics are found. Projection effects could also hamper the detection of a shock.
If the merging axis is inclined along the line of sight, the projected surface brightness map will not highlight any noticeable feature, and any shock surface will be smeared out in the surface brightness profiles.
\CORRS{However, geometrical considerations would suggest that in double-relic systems it is likely that the merging axis is not significantly inclined along the line of sight.}
Finally, the shock surface thickness may not correspond to a sharp edge but to a much wider and  more complex structure that in combination with projection effects could be difficult to detect even with a deep observation \CORRS{\citep[e.g.][]{vanWeeren2010}}. A deeper X-ray observation allowing the extraction of temperature profiles and surface brightness profiles with better photon statistics will allow us to discuss quantitatively these possibilities.}

\subsection{Double relic scaling relations}\label{sec:discussion:scaling}

\begin{figure}[!t]
    \centering
    \begin{subfigure}{1\linewidth}
    \includegraphics[width=1\linewidth]{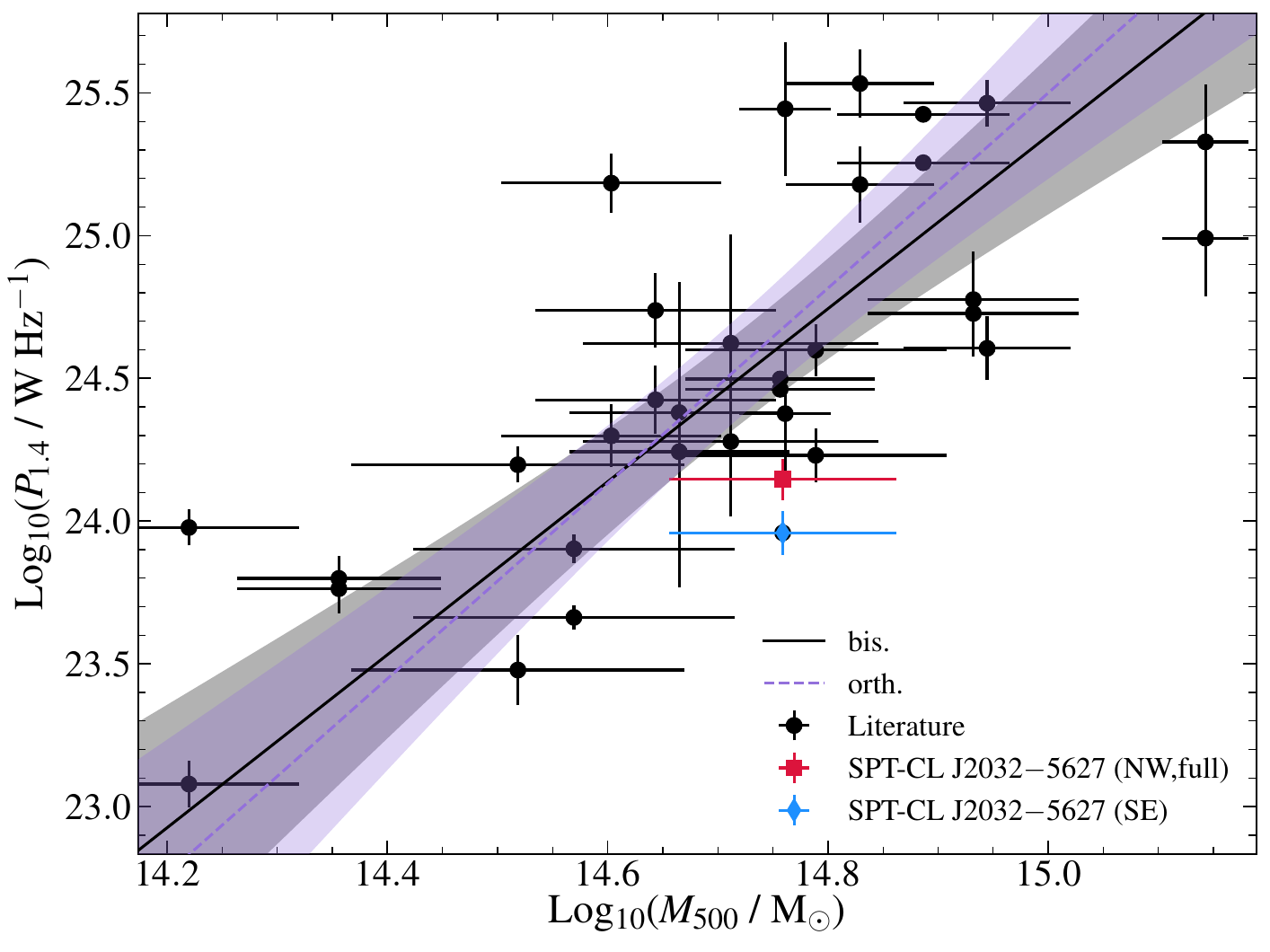}
    \caption{\label{fig:pm:double}}
    \end{subfigure}\\%
    \begin{subfigure}{1\linewidth}
    \includegraphics[width=1\linewidth]{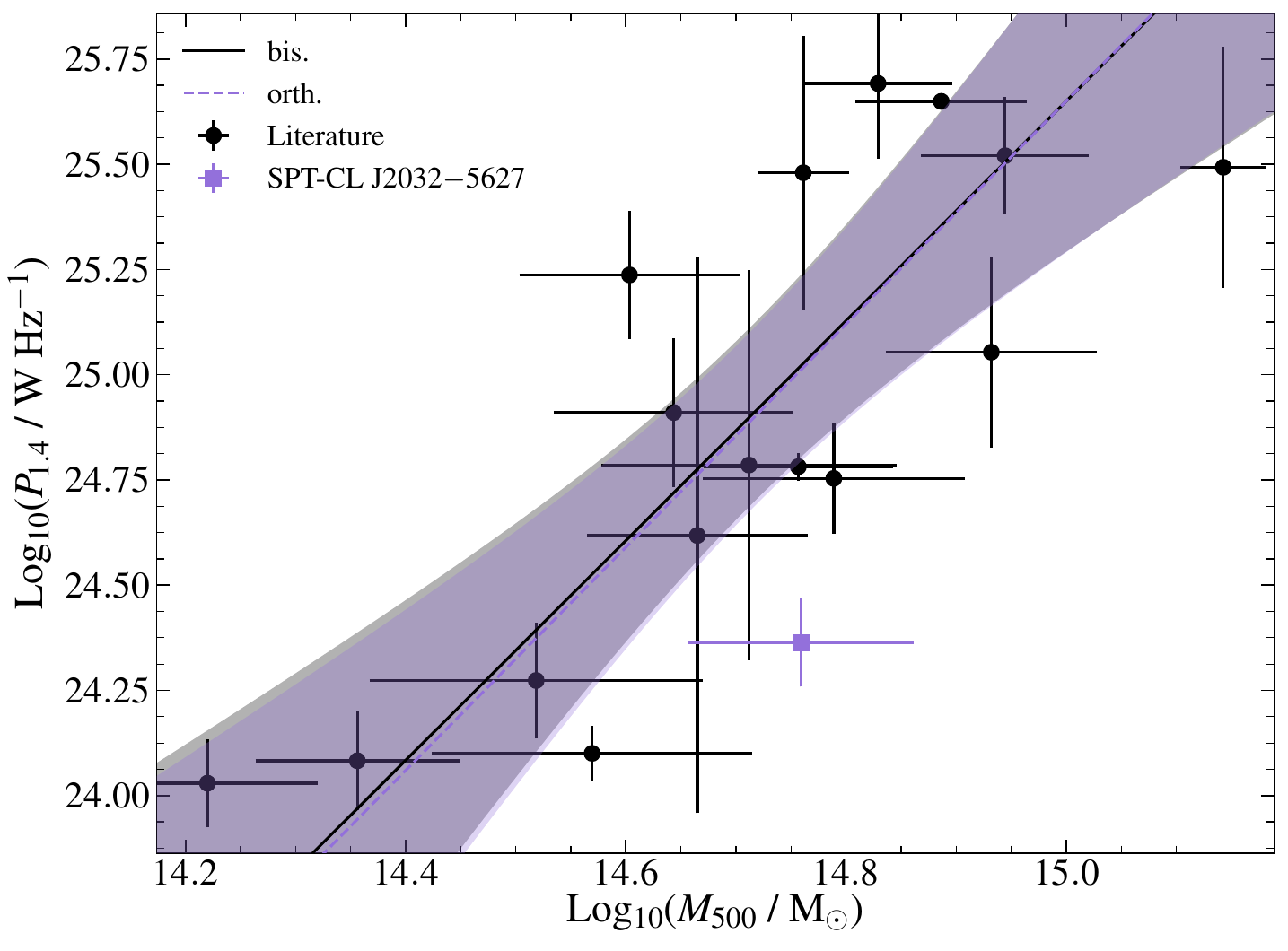}
    \caption{\label{fig:pm:sum}}
    \end{subfigure}%
    \caption{\label{fig:pm} $P$--$M$ for cluster systems hosting double relics (see Table \ref{table:relics}). \subref{fig:pm:double} Counting relics as individual sources. \subref{fig:pm:sum} Sum of power from both relics. We show the fitting results for the bisector (black-solid, grey-shaded) and orthogonal (purple-dashed, purple-shaded) methods, with shaded 95~per cent confidence intervals. Table \ref{tab:pm} presents the best fit values for the relation $\log\left(P_{1.4}\right) = A\log\left(M_{500}\right) + B$.}
\end{figure}

\begin{table}
    \centering
    
    \begin{threeparttable}
    \caption{\label{tab:pm}Best-fit parameters to $P$--$M$ relation for dRS sources list in Table \ref{table:relics}.}
    \begin{tabular}{c c c c}\toprule
    Method & $A$ & $B$ & $\sigma_\mathrm{raw}$ \\\midrule
    \multicolumn{4}{c}{Individual dRS} \\\midrule
$ P_{1.4}\vert M_{500}$ & $2.67 \pm 0.48$ & $-14.70 \pm 7.12$ & 0.47 \\
$M_{500} \vert P_{1.4}$ & $3.50 \pm 0.58$ & $-27.00 \pm 8.48$ & 0.57 \\
bis.  & $3.03 \pm 0.47$ & $-20.10 \pm 6.90$ & 0.51 \\
orth. & $3.42 \pm 0.57$ & $-25.80 \pm 8.38$ & 0.56 \\\midrule

    \multicolumn{4}{c}{Summed dRS} \\\midrule
$ P_{1.4}\vert M_{500}$ & $2.55 \pm 0.55$ & $-12.70 \pm 8.02$ & 0.42 \\
$M_{500} \vert P_{1.4}$ & $2.66 \pm 0.65$ & $-14.30 \pm 9.53$ & 0.42 \\
bis.  & $2.61 \pm 0.54$ & $-13.50 \pm 7.92$ & 0.41 \\
orth. & $2.65 \pm 0.62$ & $-14.10 \pm 9.06$ & 0.41 \\\bottomrule

    \end{tabular}
    \begin{tablenotes}[flushleft]
    \footnotesize \item \emph{Notes.} The $P$--$M$ relation is fit in the form $\log\left(P_{1.4}\right) = A\log\left(M_{500}\right) + B$.
    \end{tablenotes}
    \end{threeparttable}
\end{table}

\defcitealias{dvb+14}{DVB14}
\citeauthor{dvb+14} (\citeyear{dvb+14}; hereafter \citetalias{dvb+14}) investigated the scaling relation of dRS 1.4-GHz power with the mass of the cluster ($P_{1.4}$--$M_{500}$). \citetalias{dvb+14} derive, from \CORRS{merger simulations of X-ray--emitting clusters \citep{Poole2006}}, a relationship of the form $P \propto M^{5/3}$ but require a number of assumptions including \CORRS{spherical symmetry} and the mass--temperature relation for \CORRS{X-ray--emitting} clusters ($T \propto M^{2/3}$; \citealt{pcab09}) which is largely based on relaxed galaxy clusters. Additionally, \citetalias{dvb+14} use a simulation of 20 \CORRS{merging clusters, simulating relics at shocks and varying the magnetic field (uniform field}, a model derived from the Coma Cluster, and an equipartition model) and find steeper relations ($P \propto M^{3.02\pm0.96}$, $P \propto M^{4.73\pm1.78}$, and $P \propto M^{6.43\pm1.84}$, respectively). The uniform magnetic field model from simulations is consistent with \CORRS{the the empirical relation they find ($P \propto M^{2.60\pm0.45}$) when the total double relic power for each system is taken into account.} \par

We update the the $P_{1.4}$--$M_{500}$ scaling relation with the current sample of dRS from \citet[][see Table \ref{tab:literature} for values and references]{vda+19} and calculate radio power from available flux densities and spectral indices. In cases where multiple flux density measurements and spectral indices are available we \CORRS{extrapolate to 1.4~GHz from the closest measurement to 1.4~GHz unless a lack of $u$--$v$ coverage is a concern. For radio relics without a measured spectral index we assume an average spectral index of $\langle \alpha \rangle = -1.2 \pm 0.3$.} We use the Bivariate Estimator for Correlated Errors and intrinsic Scatter (BCES) method \citep{ab96} \footnote{\url{http://www.astro.wisc.edu/~mab/archive/stats/stats.html}} to determine the best-fit parameters to the scaling relation of the form $\log\left(P_{1.4}\right) = A\log\left(M_{500}\right) + B$. The BCES method can be performed assuming either $M_{500}$ or $P_{1.4}$ as the independent variable ($P_{1.4}\vert M_{500}$ and $M_{500} \vert P_{1.4}$, respectively), with a bisector method, or with an orthogonal method that minimises the squared orthogonal distances. While one might expect that radio relic power is dependent on cluster mass, given the large, equivalent uncertainties on each quantity we present all BCES methods for ease of comparison to other works (e.g. \citetalias{dvb+14} use bisector). Table \ref{tab:pm} lists the best-fit values, and Fig. \ref{fig:pm:double} shows the relation treating each relic in the dRS system as individual objects, and Fig. \ref{fig:pm:sum} shows the same relation but treating the whole dRS system as a single object with best-fit bisector and orthogonal lines. The summed power case shows both consistency with \citetalias{dvb+14} and between the bisector and orthogonal methods. We note that while consistent with the general scatter in the dRS systems, \spt~is a lower-brightness dRS system and falls below the best-fit scaling relation.

\subsection{Telescope capabilities in the era of Square Kilometre Array precursors}\label{sec:discussion:capabilities}

\begin{figure}
    \centering
    \includegraphics[width=1\linewidth]{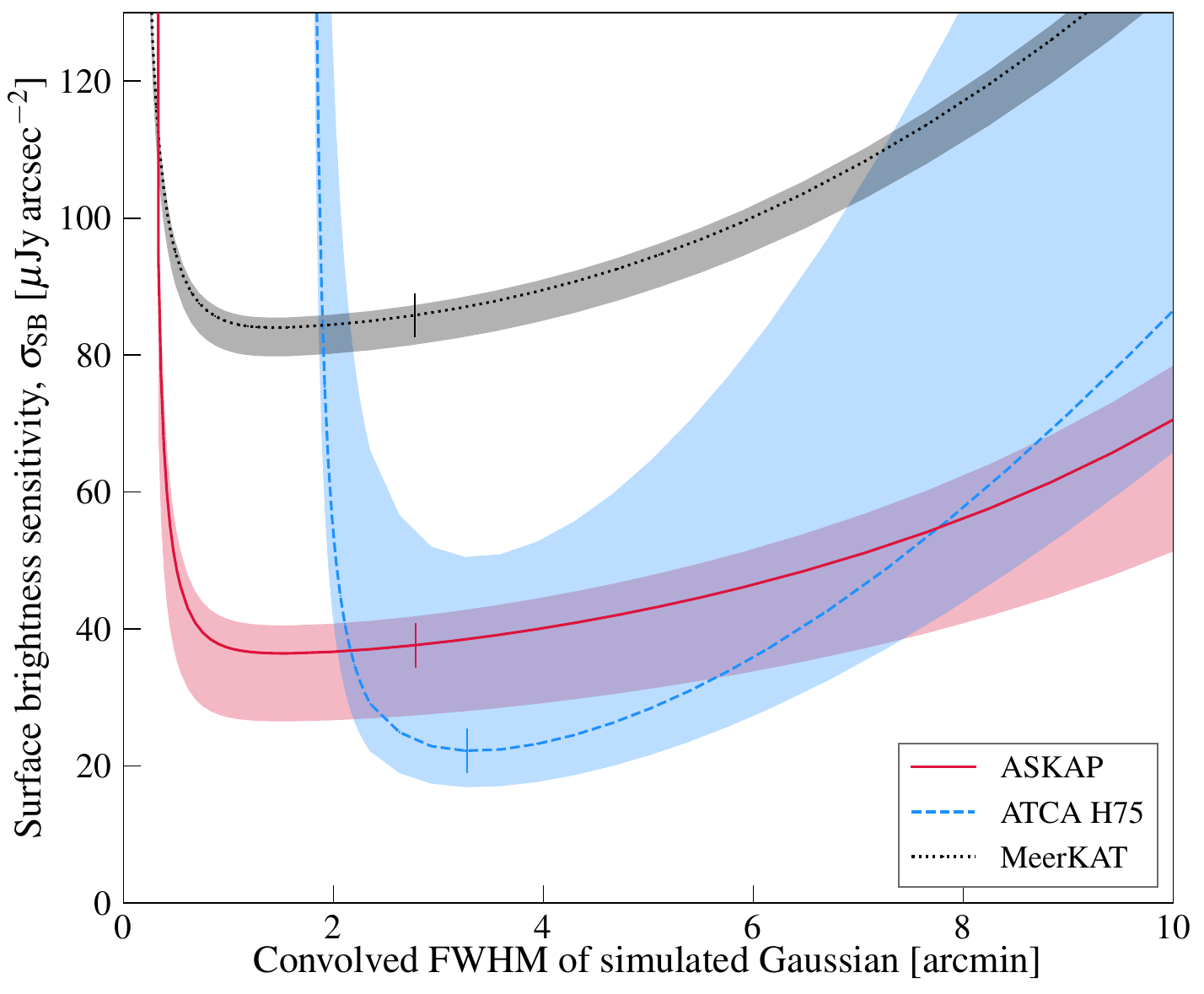}
    \caption{\label{fig:sb} Surface brightness sensitivity comparison between \CORRS{ASKAP, ATCA H75, and a MeerKAT observation as described in the text}. \CORRStwo{The data are scaled to \CORRS{1.4}~GHz using \CORRS{$\alpha=-1.2$} and the shaded regions indicate a range of spectral indices}, \CORRS{$-1.8 \leq \alpha \leq -1.0$}. The angular scale (FWHM) has been convolved with the \CORRSthree{major axis of the} restoring beam \CORRSthree{(ASKAP: 16.2~arcsec; ATCA H75: 105~arcsec; MeerKAT: 10.9~arcsec, assuming circular beams)}. The vertical notches on each dataset indicate the \CORRSthree{angular size of the SE relic after convolution with the major axis of the restoring beam}.}
\end{figure}

\CORRS{Two of the greatest outstanding mysteries associated with radio relic and halo generation are the source of the seed electron populations and the precise acceleration mechanisms at work. Previously, the rarity of radio relics and the lack of detailed observations over a range of frequencies has hampered efforts to understand the source generation mechanism.} Radio relics are not generally detected above 4 GHz \citep[see e.g.][]{vda+19}, \CORRS{and even dedicated attempts to \CORRSthree{detect} such emission in some of the brightest known relics has been unsuccessful with the previous generation of telescopes \citep[e.g. 4.8 GHz observations of the northwest relic in A3667 by][]{mj-h}. Here }however, given the $u$--$v$ coverage offered by the compact H75 configuration of the ATCA we are able to make a detection \CORRS{prompting us to consider the likelihood of similar high frequency detections in the era of next generation telescopes, particularly precursors to the Square Kilometre Array (SKA)}. 

\subsubsection{ASKAP and the ATCA H75 array}

We compare the surface brightness sensitivity of the ATCA 5.5-GHz and ASKAP observations \CORRS{(observation details in Table \ref{tab:obs})} {by simulating for each observation a range of circular Gaussian sources of varying FWHM but constant peak, $P$, in Jy\,deg$^{-2}$. The various Gaussians are \CORRSthree{each simulated and} imaged \CORRSthree{separately} with \texttt{wsclean} \CORRSthree{using the otherwise empty} \verb|MODEL_DATA| \CORRSthree{column of the respective datasets.} \CORRSthree{We use the same imaging parameters as in the original robust $+0.25$ (ASKAP) and naturally-weighted (ATCA H75) images. T}he peak surface brightness, $S_\text{peak}$, in Jy\,beam$^{-1}$ is then measured from the map without CLEANing. For a $3\sigma_\text{rms}$ detection the surface brightness sensitivity can be estimated via $\sigma_\text{SB} = 3\sigma_\text{rms} \left(P/S_\text{peak}\right)$  \citep[see Section 4 of][for further details]{Hodgson2020}.} \par

To compare ASKAP to ATCA we rescale the measured peak \CORRStwo{surface brightness} \CORRS{to 1.4~GHz} using \CORRS{$\alpha = -1.2$} but also compare a range from \CORRS{$-1.8 \leq \alpha \leq -1.0$.} Fig. \ref{fig:sb} shows \CORRStwo{$\sigma_\text{SB}$} for these datasets \CORRStwo{over a range of convolved FWHM between 0 and 10~arcmin.} \CORRSthree{While the same Gaussian sources are simulated for each dataset, after convolution with the PSF of each observation the Gaussian sources in the ATCA H75 images have larger FWHM.} Immediately we can see that the H75 data, even for such a short observation, provides good surface brightness sensitivity largely due to the size of the PSF without antenna 6. \CORRS{Additionally, the \CORRSthree{angular size of the SE relic in the H75 images} occurs at the most sensitive scale for the observation.} One thing to note here is that the primary beam FWHM of the ATCA at 5.5~GHz is $\sim10$~arcmin, which limits the angular scale sensitivity without mosaicking. 

\subsubsection{ASKAP and MeerKAT within the context of all-sky surveys}

\CORRS{ATCA is limited in its extended source surveying capability by the small FoV and the limited baselines any single array configuration provides. ASKAP surveys such as EMU, with 10-hr observations covering $\sim36$~deg$^{2}$ and excellent $u$--$v$ coverage (see Fig. \ref{fig:uv}), are beginning to provide the much-needed increase to numbers of diffuse cluster sources (e.g. \citealt{HyeongHan2020,Wilber2020b}, Duchesne et al. in prep., this work). Recently, MeerKAT \footnote{Karoo Array Telescope.} \citep[]{Jonas2009,Jonas2016} \footnote{See \citealt{Mauch2020} for further technical details.} with an increase in overall sensitivity compared to ASKAP---largely due to an increased instantaneous bandwidth and its 64 antennas---is providing excellent observations of faint synchrotron structures \citep[e.g. filaments in radio galaxies;][]{Ramatsoku2020}, suggesting MeerKAT will be a powerful instrument for detailed observations of diffuse cluster emission at frequencies on the order of 1~GHz.}

\begin{table*}[!t]
    \centering
    \begin{threeparttable}
    \caption{\CORRStwo{Details of the simulated MeerKAT data, using a robust 0.0 image weighting.} \label{tab:meerkat}}
    \begin{tabular}{c c c c c c c c c}
    \toprule
    $\tau$ \tabft{\ref{tab:meerkat:sens}} & $\Delta\nu$ & $\nu_\text{c}$ & SEFD \tabft{\ref{tab:meerkat:sens}} \tabft{\ref{tab:meerkat:sefd}} & $N_\text{ants}$ \tabft{\ref{tab:meerkat:nants}} & $\sigma_\text{rms}$ \tabft{\ref{tab:meerkat:sens}} & $\theta_\text{minor}$ \tabft{\ref{tab:meerkat:sens}} \tabft{\ref{tab:meerkat:minor}}  \\
    (min) & (GHz) & (GHz) & (Jy) & & ($\mu$Jy\,beam$^{-1}$) & ($\arcsec$)\\\midrule
     16.7 & 0.856 & 1.284 & 420 & 64 & $14.1$ & 10.9 \\\bottomrule
    \end{tabular}
    \begin{tablenotes}[flushleft]
    \footnotesize \item \textit{Notes.} \CORRSthree{\ft{tab:meerkat:sens} Value from the MeerKAT sensitivity calculator as described in the text.} Columns are similar to those in Table \ref{tab:obs} with the addition of: \ft{tab:meerkat:sefd} Average system equivalent flux density used for scaling the rms noise from observed values. \ft{tab:meerkat:nants} \CORRSthree{Note the default for the sensitivity calculator is $N_\text{ants} = 60$.} \ft{tab:meerkat:minor} Size of the minor axis for the synthesized beam.
    \end{tablenotes}
    \end{threeparttable}
\end{table*}
\setcounter{ft}{0}

\CORRStwo{While MeerKAT is able to provide deeper imaging than ASKAP, its single $\sim 1$~deg instantaneous FoV limits its ability to cover the sky at a similar rate to ASKAP. At the same surveying rate as ASKAP, MeerKAT L-band would require integration times of $\sim 16.7$~min per pointing\CORRSthree{ assuming individual pointings match the present ASKAP PAF footprint which includes a 0.9~deg separation between the 36 individual primary beams. With a robust 0.0 weighting} images would reach an rms noise of $\sim 14.1$~$\mu$Jy\,beam$^{-1}$ \footnote{\CORRStwo{For MeerKAT sensitivity calculations, see the online sensitivity calculator provided by the South African Radio Astronomy Observatory, based on real observations: \url{https://archive-gw-1.kat.ac.za/public/tools/continuum_sensitivity_calculator.html}}}, close to the lower limit expected from the EMU survey. Assuming this integration time \CORRSthree{we use the \texttt{simms} \footnote{\url{https://github.com/ratt-ru/simms}} package to} simulate an observation of a source transiting through zenith with a 1.284~GHz central frequency over the full bandwidth available. Additional details of the simulated observation are shown in Table \ref{tab:meerkat}. \CORRSthree{Note that we used the full 64 antennas for sensitivity calculation and data simulation rather than the 60 typically used.}} \par
\CORRStwo{We repeat the simulation of Gaussian sources of varying FWHM as with the ASKAP and ATCA data, and show on Fig. \ref{fig:sb} the equivalent surface brightness sensitivity limits \CORRSthree{for robust 0.0 imaging}.} This illustrates that an equivalent survey would not match ASKAP's surface brightness sensitivity for large-scale structures. \CORRSfour{We note that when $S_\text{peak}$ is measured from the CLEANed MeerKAT images this results in a decrease in $S_\text{peak}$ which increases $\sigma_\text{SB}$ by up to $\sim 30$\%, dependent on simulated Gaussian FWHM. Comparatively, $\sigma_\text{SB}$ \textit{decreases} for the ATCA H75 maps (by up to $\sim50$\%) and slightly increases for ASKAP (by up to $\sim6$\%), which does not appreciably change the conclusions drawn from Fig.~\ref{fig:sb}. An additional note} here is that changing imaging weighting or adding additional tapering would increase the sensitivity to large-scale structures, though a general all-sky survey would likely optimise for point source sensitivity. This difference suggests that ASKAP will be be an excellent survey instrument for uncovering new diffuse sources \CORRSthree{within (and outside of)} galaxy clusters, while MeerKAT and the ATCA will remain excellent complementary instruments for deeper follow-up observations at frequencies of 1~GHz and higher.

\section{Summary}
Using recently released ASKAP observations we have identified a new radio relic in the cluster \spt, with a secondary relic on the opposite side of the cluster. The radio relics are detected in both the new ASKAP data at $\sim900$~MHz and archival ATCA 5.5-GHz observations. The relics have power law spectra between 800--5500~MHz, with \alphaS~and \alphaN, consistent with many examples of radio relics. The relic properties are largely consistent with the established relic population, though they lie slightly below the $P$--$M$ scaling relation for double relic systems.  

The cluster itself is morphologically disturbed, as shown by the X-ray emission from \CORRS{the} ICM as seen by \xmm. Though no shocks are detected in the X-ray surface brightness profiles, a temperature map reveals \CORRS{a potential cold front preceding the SE relic sources.} A lack of detectable shock at the radio relic locations may be due \CORRS{to a complex shock structure, perhaps formed of multiple shocks} along the line of sight, \CORRS{or insufficient depth of the available \xmm\ data}.

\CORRS{Despite radio relics featuring steep radio spectra, ASKAP surveys such as EMU will uncover a heretofore unseen radio relic population in the Southern Sky, at this surface-brightness sensitivity with many cluster systems predicted to host such objects at low power \citep[e.g.][]{Nuza2012}.}

\CORRS{Such a sample of radio relics could be followed up by deep MeerKAT,  ASKAP and/or even ATCA observations, utilising the full frequency range provided by the instruments (either instantaneously with MeerKAT or as multiple observations across the band with ASKAP), providing wide-frequency information vital to finally understanding the acceleration mechanisms at work in clusters to generate these sources.}

\begin{acknowledgements}

The authors would like to thank the anonymous referee for their comments and suggestions that have helped to improve this manuscript. SWD acknowledges an Australian Government Research Training Program scholarship administered through Curtin University. IB acknowledges financial contribution from contract ASI-INAF n. 2017-14-H.0. GWP acknowledges the French space agency, CNES.
The Australian SKA Pathfinder is part of the Australia Telescope National Facility which is managed by CSIRO. Operation of ASKAP is funded by the Australian Government with support from the National Collaborative Research Infrastructure Strategy. ASKAP uses the resources of the Pawsey Supercomputing Centre. Establishment of ASKAP, the Murchison Radio-astronomy Observatory and the Pawsey Supercomputing Centre are initiatives of the Australian Government, with support from the Government of Western Australia and the Science and Industry Endowment Fund. We acknowledge the Wajarri Yamatji people as the traditional owners of the Observatory site. 
We acknowledge the Pawsey Supercomputing Centre which is supported by the Western Australian and Australian Governments. This paper includes archived data obtained through the Australia Telescope Online Archive (\url{http://atoa.atnf.csiro.au/}). 

This research made use of a number of \texttt{python} packages: \texttt{aplpy} \citep{Robitaille2012}, \texttt{astropy} \citep{astropy:2013,astropy:2018}, \texttt{matplotlib} \citep{Hunter2007}, \texttt{numpy} \citep{Numpy2011}, \texttt{scipy} \citep{Jones2001}, and \texttt{cmasher} \citep{cmasher}. \par
This research has made use of the NASA/IPAC Extragalactic Database (NED), which is operated by the Jet Propulsion Laboratory, California Institute of Technology, under contract with the National Aeronautics and Space Administration. This work makes use of the \texttt{cubehelix} family of colourmaps \citep{cubehelix}. \par
This project used public archival data from the Dark Energy Survey (DES). Funding for the DES Projects has been provided by the U.S. Department of Energy, the U.S. National Science Foundation, the Ministry of Science and Education of Spain, the Science and Technology FacilitiesCouncil of the United Kingdom, the Higher Education Funding Council for England, the National Center for Supercomputing Applications at the University of Illinois at Urbana-Champaign, the Kavli Institute of Cosmological Physics at the University of Chicago, the Center for Cosmology and Astro-Particle Physics at the Ohio State University, the Mitchell Institute for Fundamental Physics and Astronomy at Texas A\&M University, Financiadora de Estudos e Projetos, Funda{\c c}{\~a}o Carlos Chagas Filho de Amparo {\`a} Pesquisa do Estado do Rio de Janeiro, Conselho Nacional de Desenvolvimento Cient{\'i}fico e Tecnol{\'o}gico and the Minist{\'e}rio da Ci{\^e}ncia, Tecnologia e Inova{\c c}{\~a}o, the Deutsche Forschungsgemeinschaft, and the Collaborating Institutions in the Dark Energy Survey.
The Collaborating Institutions are Argonne National Laboratory, the University of California at Santa Cruz, the University of Cambridge, Centro de Investigaciones Energ{\'e}ticas, Medioambientales y Tecnol{\'o}gicas-Madrid, the University of Chicago, University College London, the DES-Brazil Consortium, the University of Edinburgh, the Eidgen{\"o}ssische Technische Hochschule (ETH) Z{\"u}rich,  Fermi National Accelerator Laboratory, the University of Illinois at Urbana-Champaign, the Institut de Ci{\`e}ncies de l'Espai (IEEC/CSIC), the Institut de F{\'i}sica d'Altes Energies, Lawrence Berkeley National Laboratory, the Ludwig-Maximilians Universit{\"a}t M{\"u}nchen and the associated Excellence Cluster Universe, the University of Michigan, the National Optical Astronomy Observatory, the University of Nottingham, The Ohio State University, the OzDES Membership Consortium, the University of Pennsylvania, the University of Portsmouth, SLAC National Accelerator Laboratory, Stanford University, the University of Sussex, and Texas A\&M University.
Based in part on observations at Cerro Tololo Inter-American Observatory, National Optical Astronomy Observatory, which is operated by the Association of Universities for Research in Astronomy (AURA) under a cooperative agreement with the National Science Foundation.

\end{acknowledgements}

\begin{appendix}

\section{Literature data}\label{sec:appendix:data}
In this section we provide the literature data used in Section \ref{sec:discussion:scaling}. For all sources we estimate power from flux density measurements and the reported spectral indices. Where multiple indices and flux density measurements are available, we choose values from the literature that are closest to 1.4~GHz (where no serious $u$--$v$ coverage problems exist). Additionally, for clusters without a measured spectra index assume a spectral index based on the mean of the sample ($\langle \alpha \rangle = -1.2 \pm 0.3$). Mass measurements are taken from SZ measurements where available (e.g. via \citealt{planck15}) or from X-ray measurements if SZ-proxy mass estimates are not available. 

\begin{table*}
    \centering
    \begin{threeparttable}
    \caption{\label{tab:literature} Clusters with dRS sources used in Section \ref{sec:discussion:scaling}.}
    \begin{tabular}{r c c c c c c}\toprule
Cluster & $z$ & $M_{500}$ & Relic No. & $\alpha$ & $P_{1.4}$ & References \\
        &     & ($\times 10^{14}$~M$_\odot$) & & & ($\times 10^{23}$~W\,Hz$^{-1}$) & \tabft{\ref{tab:references}} \\\midrule
\multirow{2}{*}{Abell 3376} & \multirow{2}{*}{0.046} & \multirow{2}{*}{$2.27\pm0.21$ \tabft{\ref{tab:psz1}}} & 1 & $-1.82\pm0.06$ & $6.3\pm0.5
$ & \tabft{\ref{tab:Kale2012}} \\
 & & & 2 & $-1.70\pm0.06$ & $5.8\pm0.5$ & \tabft{\ref{tab:Kale2012}} \\
\multirow{2}{*}{Abell 3667} & \multirow{2}{*}{0.056} & \multirow{2}{*}{$5.77\pm0.24$ \tabft{\ref{tab:psz1}}} & 1 & $-0.90\pm0.10$ & $23.8\pm5.
4$ & \tabft{\ref{tab:Hindson2014}} \\
 & & & 2 & $-0.90\pm0.10$ & $278.0\pm65.0$ & \tabft{\ref{tab:Hindson2014}} \\
\multirow{2}{*}{Abell 3365} & \multirow{2}{*}{0.093} & \multirow{2}{*}{$1.66\pm0.17$ \tabft{\ref{tab:mcxc}}} & 1 & $-1.20\pm0.30$ & $9.5\pm0.6
$ & \tabft{\ref{tab:vanWeeren2011a}}/- \\
 & & & 2 & $-1.20\pm0.30$ & $1.2\pm0.1$ & \tabft{\ref{tab:vanWeeren2011a}}/- \\
\multirow{2}{*}{ZwCl 0008.8$+$5215} & \multirow{2}{*}{0.103} & \multirow{2}{*}{$3.30\pm0.50$ \tabft{\ref{tab:psz1}}} & 1 & $-1.59\pm0.06$ & $1
5.8\pm1.0$ & \tabft{\ref{tab:vanWeeren2011b}} \\
 & & & 2 & $-1.49\pm0.12$ & $3.0\pm0.4$ & \tabft{\ref{tab:vanWeeren2011b}} \\
\multirow{2}{*}{Abell 1240} & \multirow{2}{*}{0.159} & \multirow{2}{*}{$3.71\pm0.54$ \tabft{\ref{tab:psz1}}} & 1 & $-1.20\pm0.10$ & $4.6\pm0.2
$ & \tabft{\ref{tab:Bonafede2009}} \\
 & & & 2 & $-1.30\pm0.20$ & $8.0\pm0.4$ & \tabft{\ref{tab:Bonafede2009}} \\
\multirow{2}{*}{Abell 2345} & \multirow{2}{*}{0.176} & \multirow{2}{*}{$5.71\pm0.49$ \tabft{\ref{tab:psz1}}} & 1 & $-1.50\pm0.10$ & $29.0\pm0.
7$ & \tabft{\ref{tab:Bonafede2009}} \\
 & & & 2 & $-1.30\pm0.10$ & $31.5\pm0.8$ & \tabft{\ref{tab:Bonafede2009}} \\
\multirow{2}{*}{CIZA J2242.8$+$5301} & \multirow{2}{*}{0.192} & \multirow{2}{*}{$4.01\pm0.40$ \tabft{\ref{tab:mcxc}}} & 1 & $-1.06\pm0.04$ & $
152.7\pm15.9$ & \tabft{\ref{tab:vanWeeren2010}} \\
 & & & 2 & $-1.29\pm0.04$ & $19.9\pm2.2$ & \tabft{\ref{tab:vanWeeren2010}}  \\
\multirow{2}{*}{RXC J1314.4$-$2515} & \multirow{2}{*}{0.244} & \multirow{2}{*}{$6.15\pm0.73$ \tabft{\ref{tab:psz1}}} & 1 & $-1.40\pm0.09$ & $1
7.0\pm1.6$ & \tabft{\ref{tab:Feretti2005}}/\tabft{\ref{tab:Venturi2007}} \\
 & & & 2 & $-1.41\pm0.09$ & $39.7\pm3.6$ &  \tabft{\ref{tab:Feretti2005}}/\tabft{\ref{tab:Venturi2007}} \\
\multirow{2}{*}{ZwCl 2341.1$+$0000} & \multirow{2}{*}{0.270} & \multirow{2}{*}{$5.15\pm0.69$ \tabft{\ref{tab:psz1}}} & 1 & $-0.49\pm0.18$ & $1
9.0\pm5.0$ & \tabft{\ref{tab:vanWeeren2009}} \\
 & & & 2 & $-0.76\pm0.17$ & $42.0\pm16.0$ & \tabft{\ref{tab:vanWeeren2009}} \\
\multirow{2}{*}{SPT-CL J2032$-$5627} & \multirow{2}{*}{0.284} & \multirow{2}{*}{$5.74\pm0.59$ \tabft{\ref{tab:psz2}}} & 1 & \alphaNonly & $
\powerN$ & this work \\
 & & & 2 & \alphaSonly & $\powerS$ & this work \\
\multirow{2}{*}{PSZ1 G096.89$+$24.17} & \multirow{2}{*}{0.300} & \multirow{2}{*}{$4.40\pm0.48$ \tabft{\ref{tab:psz1}}} & 1 & $-1.20\pm0.30$ & 
$26.6\pm3.2$ & \tabft{\ref{tab:deGasperin2014}}/- \\
 & & & 2 & $-1.20\pm0.30$ & $54.7\pm7.1$ & \tabft{\ref{tab:deGasperin2014}}/- \\
\multirow{2}{*}{PSZ1 G108.18$-$11.53} & \multirow{2}{*}{0.335} & \multirow{2}{*}{$7.70\pm0.60$ \tabft{\ref{tab:psz1}}} & 1 & $-1.25\pm0.02$ & 
$266.0\pm1.9$ & \tabft{\ref{tab:deGasperin2015}} \\
 & & & 2 & $-1.28\pm0.02$ & $180.0\pm1.6$ & \tabft{\ref{tab:deGasperin2015}}  \\
\multirow{2}{*}{MACS J1752.0$+$4440} & \multirow{2}{*}{0.367} & \multirow{2}{*}{$6.75\pm0.45$  \tabft{\ref{tab:psz2}}} & 1 & $-1.21\pm0.06$ & $
341.0\pm41.0$ & \tabft{\ref{tab:Bonafede2012}} \\
 & & & 2 & $-1.12\pm0.07$ & $151.0\pm20.0$ & \tabft{\ref{tab:Bonafede2012}} \\
\multirow{2}{*}{PSZ1 G287.0$+$32.9} & \multirow{2}{*}{0.390} & \multirow{2}{*}{$13.89\pm0.54$  \tabft{\ref{tab:psz1}}} & 1 & $-1.36\pm0.21$ & $
213.0\pm43.0$ & \tabft{\ref{tab:Bonafede2014}} \\
 & & & 2 & $-1.33\pm0.21$ & $98.0\pm20.0$ & \tabft{\ref{tab:Bonafede2014}} \\
\multirow{2}{*}{MACS J1149.5$+$2223} & \multirow{2}{*}{0.544} & \multirow{2}{*}{$8.55\pm0.82$  \tabft{\ref{tab:psz1}}} & 1 & $-1.15\pm0.08$ & $
53.4\pm8.0$ & \tabft{\ref{tab:Bonafede2012}} \\
 & & & 2 & $-0.75\pm0.08$ & $59.7\pm10.1$ & \tabft{\ref{tab:Bonafede2012}} \\
\multirow{2}{*}{MACS J0025.4$-$1222} & \multirow{2}{*}{0.586} & \multirow{2}{*}{$4.62\pm0.46$  \tabft{\ref{tab:mcxc}}} & 1 & $-1.20\pm0.30$ & $
17.5\pm8.3$ & \tabft{\ref{tab:Riseley2017}}/- \\
 & & & 2 & $-1.20\pm0.30$ & $24.0\pm11.0$ & \tabft{\ref{tab:Riseley2017}}/-  \\
\multirow{2}{*}{ACT-CL J0102$-$4915} & \multirow{2}{*}{0.870} & \multirow{2}{*}{$8.80\pm0.67$ \tabft{\ref{tab:psz1}}} & 1 & $-1.19\pm0.09$ & $
291.0\pm24.0$ & \tabft{\ref{tab:Lindner2014}} \\
 & & & 2 & $-1.40\pm0.10$ & $40.3\pm4.5$ & \tabft{\ref{tab:Lindner2014}} \\
\bottomrule
\end{tabular}
\begin{tablenotes}[flushleft]
\footnotesize \item \emph{Notes.} \ft{tab:references} flux density/$\alpha$. ``-'' $= \langle \alpha \rangle$. \textit{References.} \ft{tab:psz1} \citet{planck15}. \ft{tab:mcxc} \citet{pap+11}; assume 10 per cent uncertainty on mass estimate. \ft{tab:psz2} \citet{planck16}. \ft{tab:Kale2012} \citet{kale_spectral_2012}. \ft{tab:Hindson2014} \citet{Hindson2014}. \ft{tab:vanWeeren2011a} \citet{vanWeeren2011a}. \ft{tab:vanWeeren2011b} \citet{vanWeeren2011b}. \ft{tab:Bonafede2009} \citet{Bonafede2009}. \ft{tab:vanWeeren2010} \citet{vanWeeren2010}. \ft{tab:Feretti2005} \citet{feretti_diffuse_2005}. \ft{tab:Venturi2007} \citet{Venturi2007}. \ft{tab:vanWeeren2009} \citet{vanWeeren2009}. \ft{tab:deGasperin2014} \citet{dvb+14}. \ft{tab:deGasperin2015} \citet{div+15}. \ft{tab:Bonafede2012} \citet{Bonafede2012}. \ft{tab:Bonafede2014} \citet{Bonafede2014}. \ft{tab:Riseley2017} \citet{riseley_diffuse_2017}. \ft{tab:Lindner2014} \citet{lindner_radio_2014}.
\end{tablenotes}
\end{threeparttable}
\end{table*}

\end{appendix}

{\footnotesize
\bibliographystyle{pasa-mnras}
\bibliography{bib_file}
}

\end{document}